\documentclass[fleqn,times,usenatbib]{mnras}

\usepackage{mathptmx}

\usepackage[figuresright]{rotating}
\usepackage{longtable}

\usepackage[T1]{fontenc}

\DeclareRobustCommand{\VAN}[3]{#2}
\let\VANthebibliography\thebibliography
\def\thebibliography{\DeclareRobustCommand{\VAN}[3]{##3}\VANthebibliography}

\usepackage{graphicx}	
\usepackage{amsmath}	
\usepackage{amssymb}	

\hypersetup{linkcolor=red,citecolor=blue,filecolor=cyan,urlcolor=magenta}

\newcommand{\angstrom}{\text{\normalfont\AA}}

\def\obj{\mbox{SRGE J170245.3+130104}}

\def\tar{J1702+1301}

\title[Is J1702+1301 a blazar?]{Is the  X-ray bright $z=5.5$ quasar \obj\ a blazar?}

\author[An et al.]{
Tao An$^{1,2}$\thanks{E-mail: antao@shao.ac.cn},
Ailing Wang$^{1,2}$,
Yuanqi Liu$^1$, 
Yulia Sotnikova$^3$,
Yingkang Zhang$^{1}$,
\newauthor 
J.N.H.S. Aditya$^{4,5}$, 
Sumit Jaiswal$^1$,
George Khorunzhev$^6$,
Baoqiang Lao$^{1,7}$,
Ruqiu Lin$^{1,2}$,
\newauthor 
Alexander Mikhailov$^{3}$,
Marat Mingaliev$^{3,8,9}$,
Timur Mufakharov$^{3,8}$,
Sergey Sazonov$^6$
\\
$^{1}$ Shanghai Astronomical Observatory, CAS, 80 Nandan Road, Shanghai 200030, China \\
$^{2}$ School of Astronomy and Space Sciences, University of Chinese Academy of Sciences, No. 19A Yuquan Road, Beijing 100049, China  \\
$^{3}$ Special Astrophysical Observatory of RAS, Nizhny Arkhyz, 369167, Russia \\
$^{4}$ Sydney Institute of Astronomy, School of Physics A28, University of Sydney, NSW 2006, Australia \\
$^{5}$ ARC Centre of Excellence for All Sky Astrophysics in 3 Dimensions (ASTRO 3D) \\
$^{6}$ Space Research Institute RAS, Moscow, 117997, Russia \\
$^{7}$ School of Physics and Astronomy, Yunnan University, Kunming, 650091, China\\
$^{8}$ Kazan Federal University, 18 Kremlyovskaya St, Kazan 420008, Russia\\
$^{9}$ Institute of Applied Astronomy RAS, St. Petersburg 191187, Russia
}

\date{Accepted XXX. Received YYY; in original form ZZZ}

\pubyear{2022}

\begin{document}
\label{firstpage}
\pagerange{\pageref{firstpage}--\pageref{lastpage}}
\maketitle

\begin{abstract}
Jets may have contributed to promoting the growth of seed black holes in the early Universe, and thus observations of radio-loud high-redshift quasars are crucial to understanding the growth and evolution of the early supermassive black holes. Here we report the radio properties of an X-ray bright $z=5.5$ quasar, SRGE J170245.3+130104 (J1702+1301). Our high-resolution radio images reveal the radio counterpart at the optical position of J1702+1301, while another radio component is also detected at $\sim$23.5\arcsec\ to the southwest. Our  analysis suggests that this southwest component is associated with a foreground galaxy at $z\approx 0.677$, which is mixed with J1702+1301 in low-frequency low-resolution radio images. After removing the contamination from this foreground source, we recalculated the radio loudness of J1702+1301 to be $R>$1100, consistent with those of blazars. J1702+1301 exhibits a flat radio spectrum ($\alpha = -0.17 \pm 0.05$, $S \propto \nu^\alpha$) between 0.15 and 5 GHz; above 5 GHz, it shows a rising spectrum shape, and the spectral index $\alpha^{8.2}_{4.7}$ appears to be correlated with the variation of the flux density: in burst states, $\alpha^{8.2}_{4.7}$ becomes larger. J1702+1301 displays distinct radio variability on timescales from weeks to years in the source's rest frame. These radio properties, including high radio loudness, rising spectrum, and rapid variability, tend to support it as a blazar. 

\end{abstract}


\section{Introduction} \label{sec:intro}

Recently, the eROSITA telescope \citep{2021A&A...647A...1P} on-board the Spectrum Roentgen Gamma \citep[SRG;][]{2021A&A...656A.132S} space observatory discovered a luminous quasar, SRGE J170245.3+130104 (J1702+1301), during its first half-year X-ray All-sky Survey \citep{2021AstL...47..123K}.  Follow-up optical spectroscopic observations conducted with the Bolshoi Teleskop Alt-azimutalnyi (BTA) 6-m telescope revealed a broad Ly$\alpha$ emission line with neutral hydrogen (H{\sc I}) absorption \citep[][]{2021AstL...47..123K}; the authors estimated a redshift of $z_{\rm em} = 5.466 \pm 0.003$.
\tar\ was detected by the eROSITA twice, on 2020 March 13--15 and 2020 September 13--14, with a nearly two-fold decrease in flux density over a six-month time span.
In the first observation, the X-ray luminosity of this quasar was $3.6 \times 10^{46}$ erg s$^{-1}$, in the energy range of 2--10 keV in the source's rest frame, making it one of the most luminous X-ray quasars at $z > 5$ \citep[][]{2019A&A...630A.118V, 2021AstL...47..123K, 2021ApJ...906..135L, 2020MNRAS.497.1842M}.   

The bolometric luminosity of \tar\ is estimated to be $L_{\rm bol} \approx 10^{47}$erg~s$^{-1}$, yielding a lower limit on its black hole mass of $>10^9 M_\odot$, assuming that the bolometric luminosity does not exceed the critical Eddington luminosity \citep{2021AstL...47..123K}. 
The bolometric luminosity and black hole mass of \tar\ are in good agreement with the $z \gtrsim 6$ quasars already discovered \citep[e.g.][]{2014ApJ...790..145D,2017ApJ...849...91M,2019ApJ...884...30W}.  
\tar\ represents the tip of the iceberg of high-$z$ extremely luminous quasars that contain supermassive black holes (SMBHs) which were formed rapidly in the early Universe and accreting at high Eddington ratios. 
The existence of such SMBHs just $\sim$1 Gyr after the Big Bang places strong constraints on the models of seed black hole formation. Jets may promote rapid accretion and thus accelerate the growth of seed black holes into SMBHs \citep{2008MNRAS.386..989J,2013MNRAS.432.2818G}, therefore studying high-$z$ jetted AGN helps to test this hypothesis.

The radio counterpart of \tar\ has been detected in the NRAO VLA Sky Survey \citep[NVSS,][]{1998AJ....115.1693C} image at 1.4 GHz, with a flux density of $\sim 28$ mJy, one of the few radio-brightest sources among the high-redshift quasars at $z \gtrsim 5$ \citep{2017FrASS...4....9P,2020A&A...635L...7B}. 
Notably, two other $z > 5$ quasars (B2 1023+25, \citealt{2012MNRAS.426L..91S}; Q0906+6930, \citealt{2004ApJ...610L...9R}), with similar X-ray fluxes and higher radio flux densities \citep{2015MNRAS.446.2921F,2017MNRAS.468...69Z}, have been identified as blazars \citep{2004ApJ...610L...9R,2012MNRAS.426L..91S,2020NatCo..11..143A}. The natural idea of finding high-redshift luminous sources from the eROSITA survey comparable to the X-ray luminosities of the two known earliest blazars is that the newly detected objects are also blazars.

\tar\ has also been detected in the Very Large Array Sky Survey (VLASS) at 3 GHz \citep{2020RNAAS...4..175G}, but not detected by the Giant Metrewave Radio Telescope (GMRT) at 150 MHz \citep{2017A&A...598A..78I}. 
Its radio loudness parameter $R$ \footnote{The conventional radio-optical radio-loudness parameter $R$ is defined as the ratio between two monochromatic fluxes $S_{\rm 5GHz}/S_{\rm 4400 \angstrom}$\citep{1989AJ.....98.1195K},  where $R<10$ for radio-quiet AGNs and $R>10$ for radio-loud AGNs.}
is preliminarily estimated to be $\sim 1200$  \citep{2021AstL...47..123K}, a typical value for low-redshift blazars but an extremely high value for high-redshift quasars \citep{2015ApJ...804..118B,2021MNRAS.508.2798S}.  
Among the active galactic nuclei (AGNs) with $z>5$, only a very small number of them, including \tar\, have $R$ exceeding $10^3$ \citep{2004ApJ...610L...9R,2018ApJ...861L..14B,2019A&A...629A..68B,2021A&A...647L..11I,2021MNRAS.508.2798S}.
In the NVSS image (resolution of 45\arcsec), the radio source remains an unresolved structure, while in the VLASS (resolution of 2.5\arcsec) image, the NVSS source is resolved into two components. Higher-resolution radio observational data are needed to characterise the radio emission of \tar.

This paper presents new radio images of \tar\ from GMRT, Murchison Widefield Array (MWA) and Australian Square Kilometre Array Pathfinder (ASKAP) observations, and flux density monitoring data from the RATAN-600 telescope. These data help us to understand more clearly the classification and radio properties of this high-$z$ X-ray bright quasar.

\section{Data and results}
\label{sec:result}

\subsection{MWA}
 
\begin{figure*}
    \centering
    \includegraphics[width=0.33\textwidth]{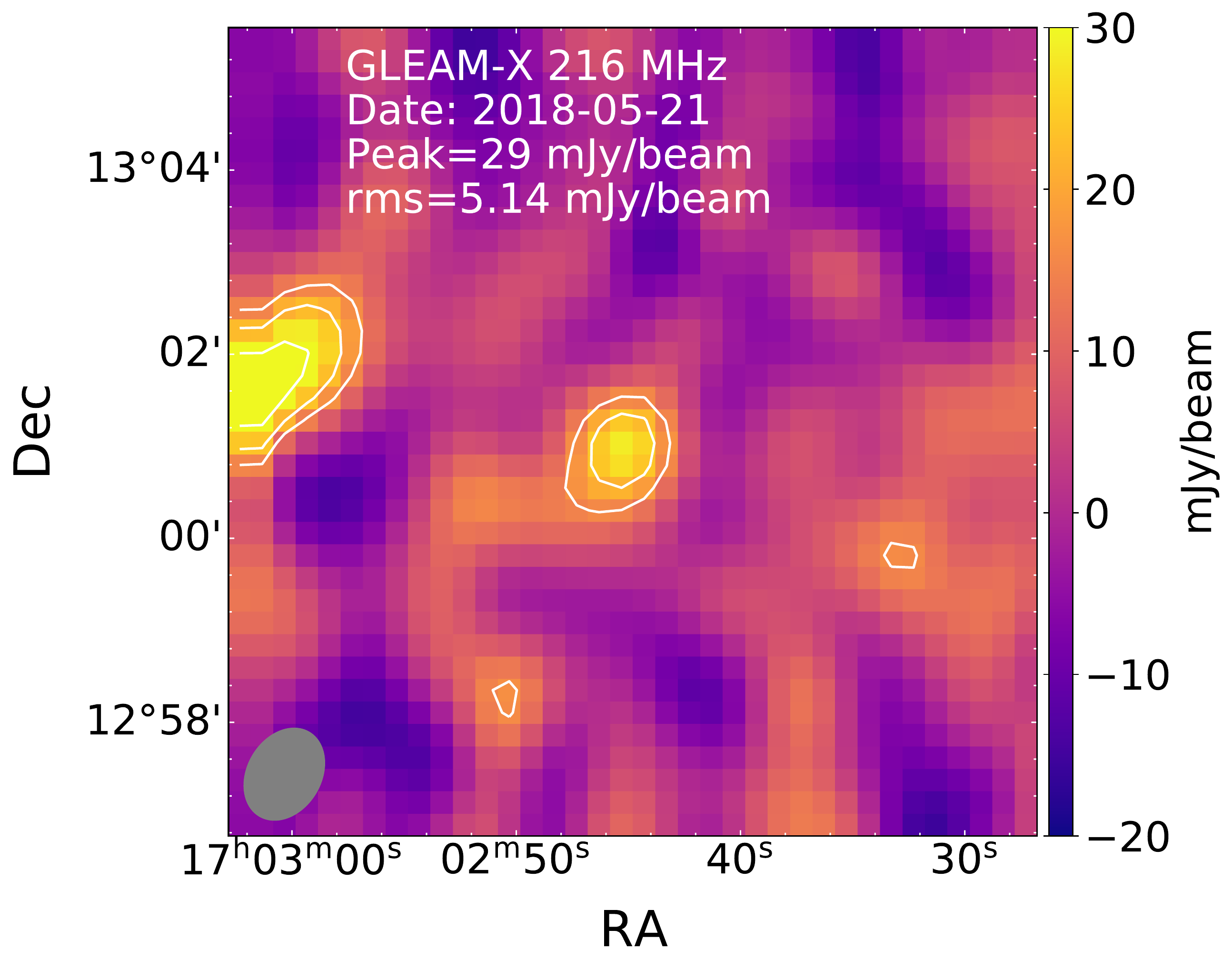}
    \includegraphics[width=0.33\textwidth]{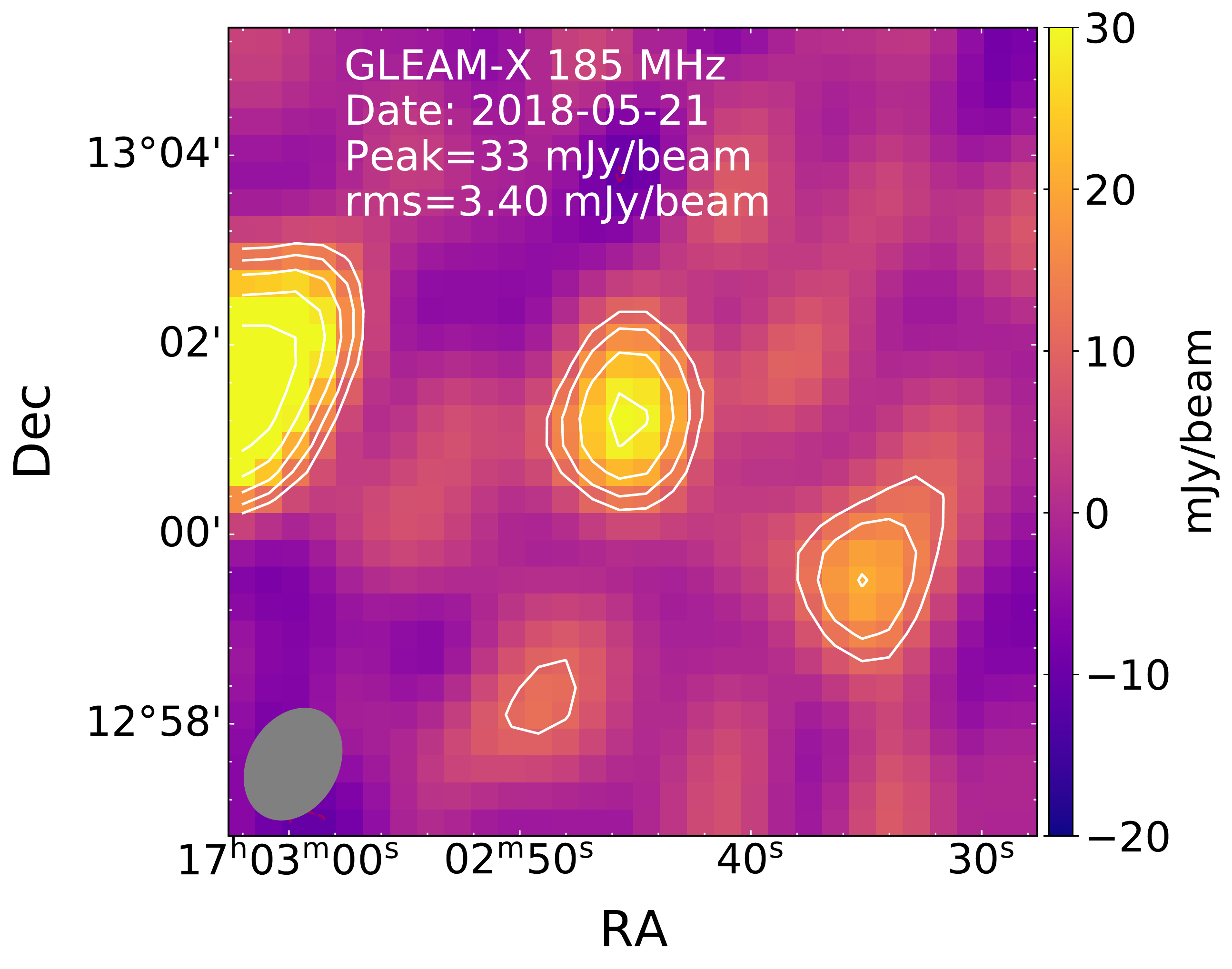}
    \includegraphics[width=0.33\textwidth]{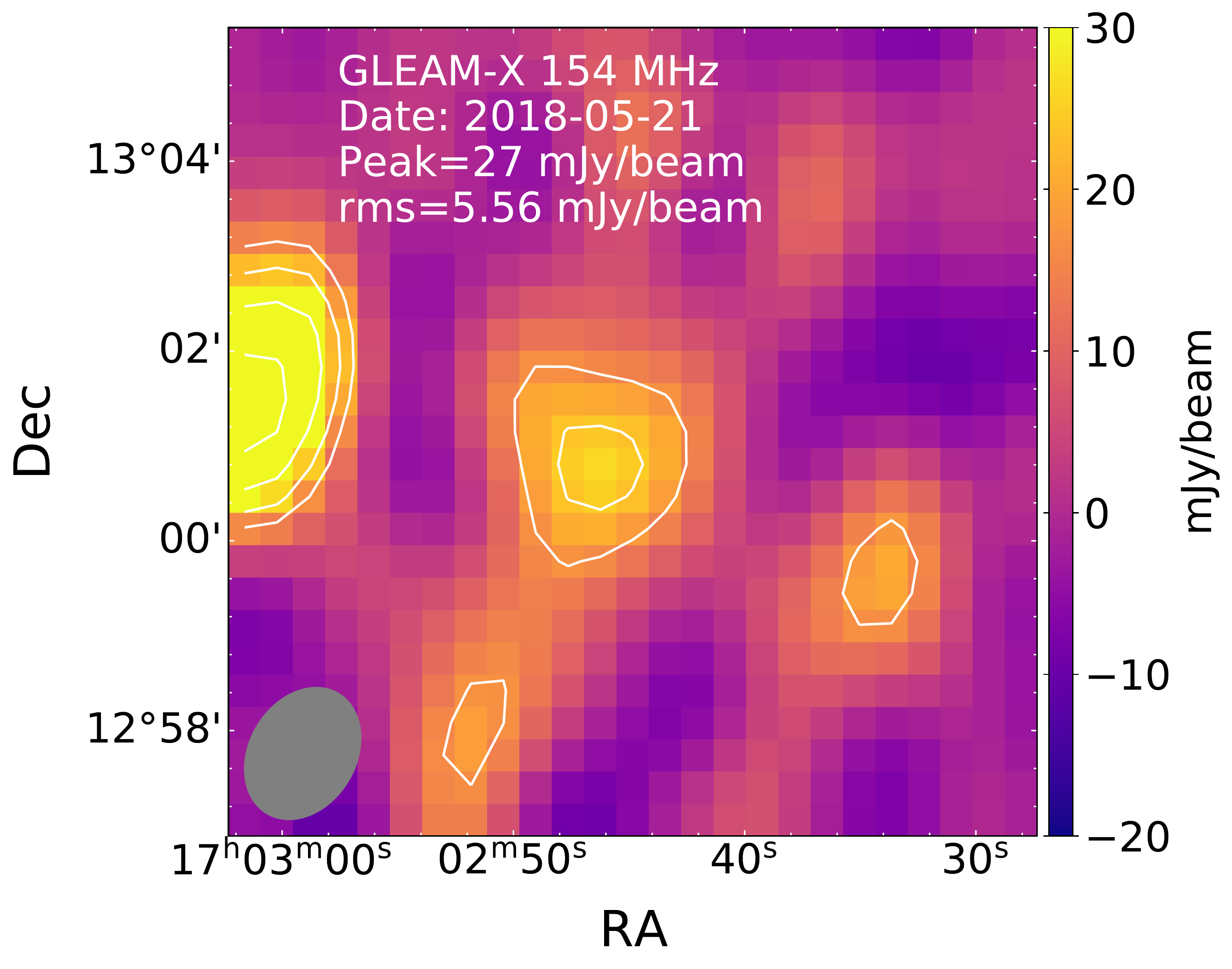}\\
    \caption{MWA images of \tar\ obtained from the GLEAM-X survey. The observation date, frequency, peak intensity, and \textit{rms} noise of the images are labelled in each panel. The images are composited from several snapshots.  We chose the middle observation date to display in the images. The restoring beam is shown as an ellipse in the bottom-left corner. The contours start at three times \textit{rms} noise and increase in a step of 2. The colour bar shows the intensity scale.}
    \label{fig:GLEAMX}
\end{figure*}

The MWA \citep{2013PASA...30....7T} is the precursor telescope of the Square Kilometre Array (SKA) low-frequency array. 
The GaLactic and Extragalactic All-sky Murchison Widefield Array (GLEAM) survey is the wide-field continuum imaging survey of the MWA \citep{2015PASA...32...25W} covering the sky south of declination +30$\degr$ over a frequency range of 72–231 MHz \citep{2017MNRAS.464.1146H}. 
GLEAM was carried out with the MWA Phase I between August 2013 and July 2015, with the longest baseline of $\sim3$ km. The highest resolution of the GLEAM images is about 2\arcmin. The  root-mean-square (\textit{rms}) noise of GLEAM images is 10 mJy beam$^{-1}$.
We examined the GLEAM archive images \footnote{The GLEAM postage stamp service: \url{http://gleam-vo.icrar.org/gleam_postage/q/form}} and found no clear detection at the location of \tar\ due to the high \textit{rms} noise. 
Higher-sensitivity and higher-resolution radio images at these low frequencies are necessary to constrain the radio spectrum of J1702+1301 below and around $\nu_{\rm rest} = 1$ GHz (the source's rest frame). 
Therefore, we collected the unreleased data observed with the MWA Phase II, consisting of an extended array of 256 tiles and with the longest baseline of up to 6 km \citep{2018PASA...35...33W}. The  GaLactic and Extragalactic All-Sky MWA-eXtended (GLEAM-X) survey based on MWA Phase II \citep{2022PASA...39...35H} has an angular resolution of $\sim45\arcsec$ at 216 MHz, with a mean image noise lower than GLEAM by a factor of $\sim7$. The GLEAM-X data were recorded in five near-contiguous channels of 30.72 MHz each. The observations were made in the snapshot mode,  each snapshot lasting 2 minutes and cycling between the five frequency channels. 
We downloaded the GLEAM-X data to the clusters of China SKA Regional Centre \citep{2019NatAs...3.1030A,2022SCPMA..6529501A} and then processed them following the pipeline developed by the GLEAM-X team 
\footnote{\url{https://github.com/tjgalvin/GLEAM-X-pipeline}}.
The final deep images shown in Fig.~\ref{fig:GLEAMX} were obtained from a combination of the astrometrically-corrected and primary-beam-corrected snapshot images observed between 2018 March 12 and 2019 May 21 through a mosaicking procedure.

\subsection{GMRT}

\begin{figure*}
    \centering
    \includegraphics[width=0.32\textwidth]{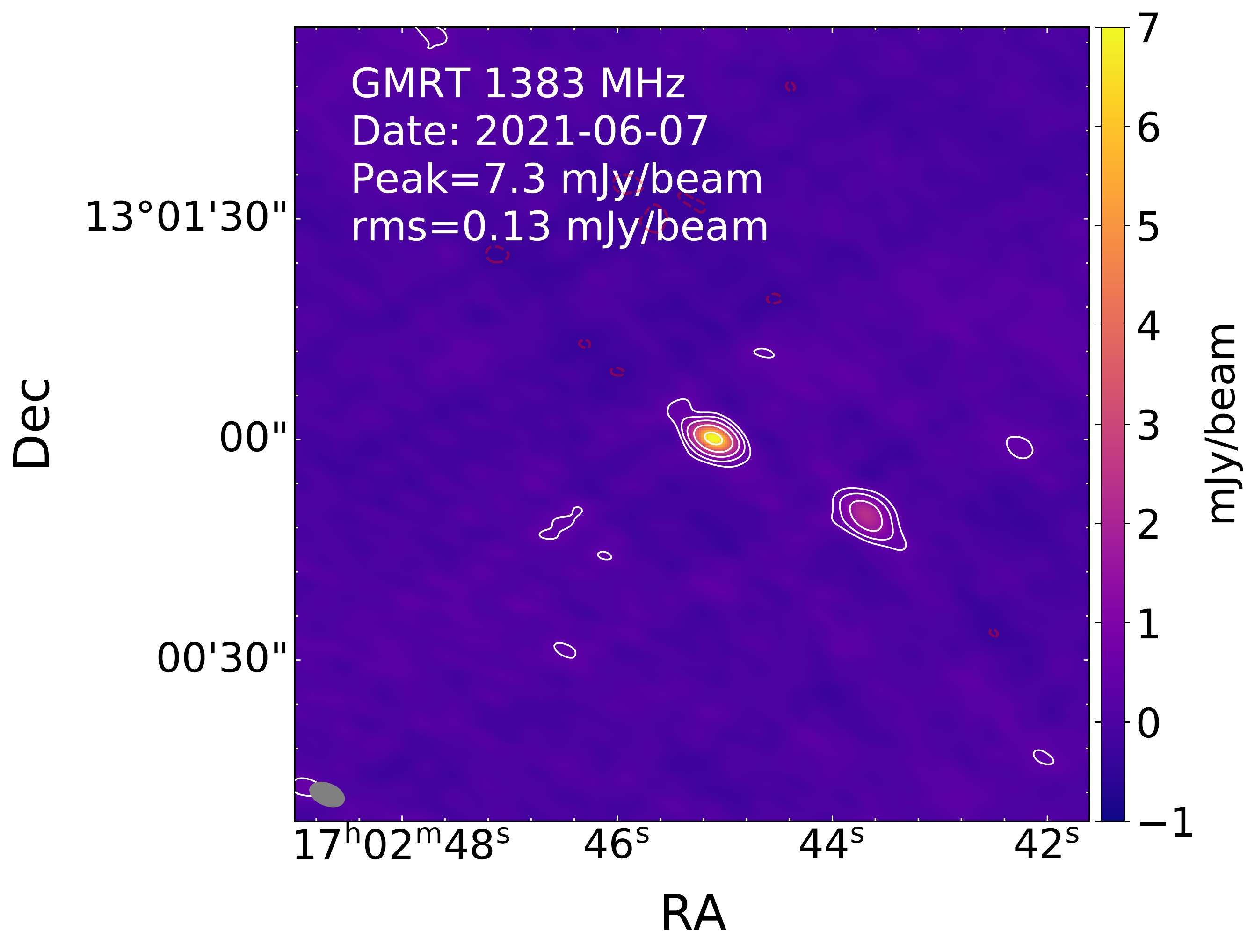}
    \includegraphics[width=0.32\textwidth]{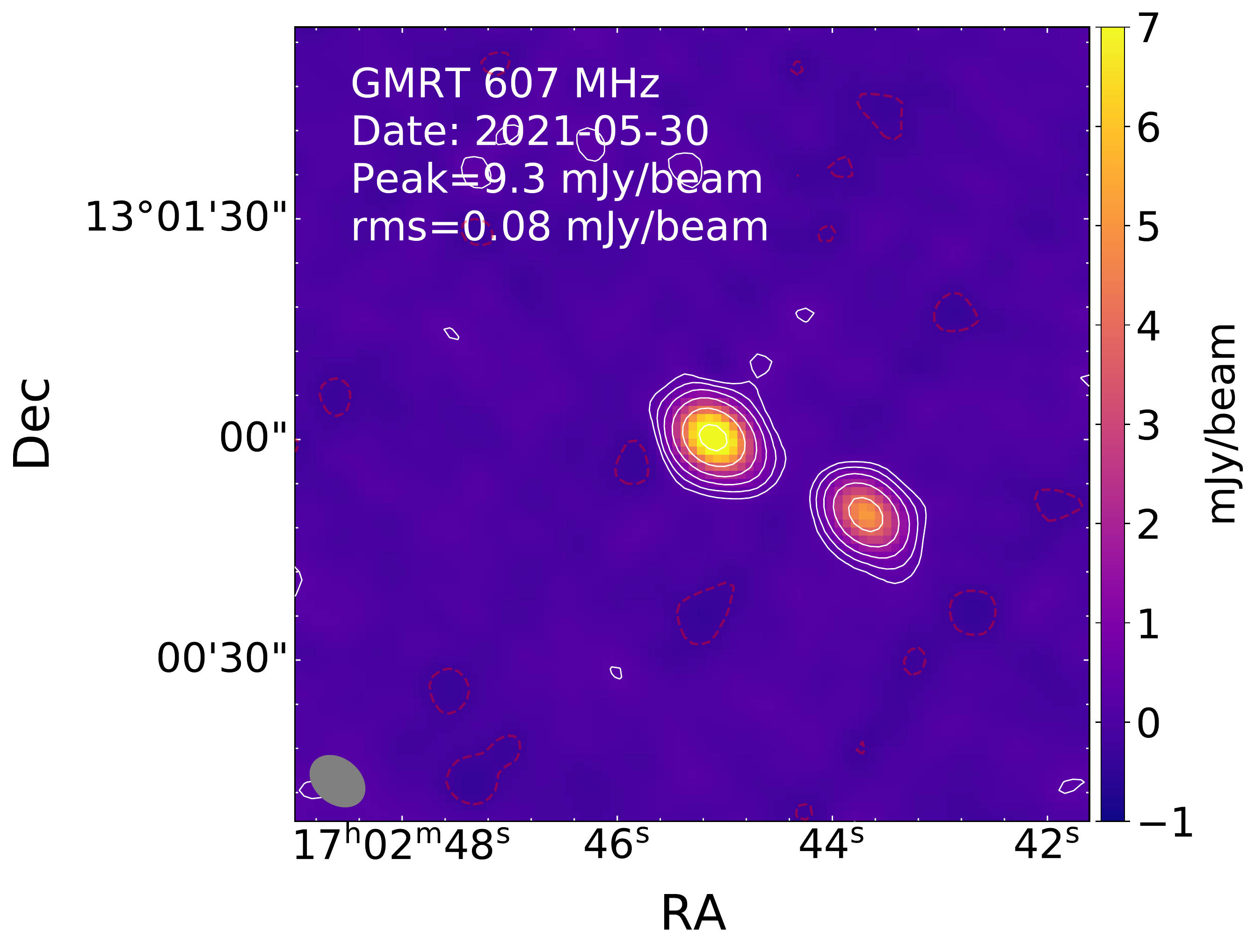}
    \includegraphics[width=0.32\textwidth]{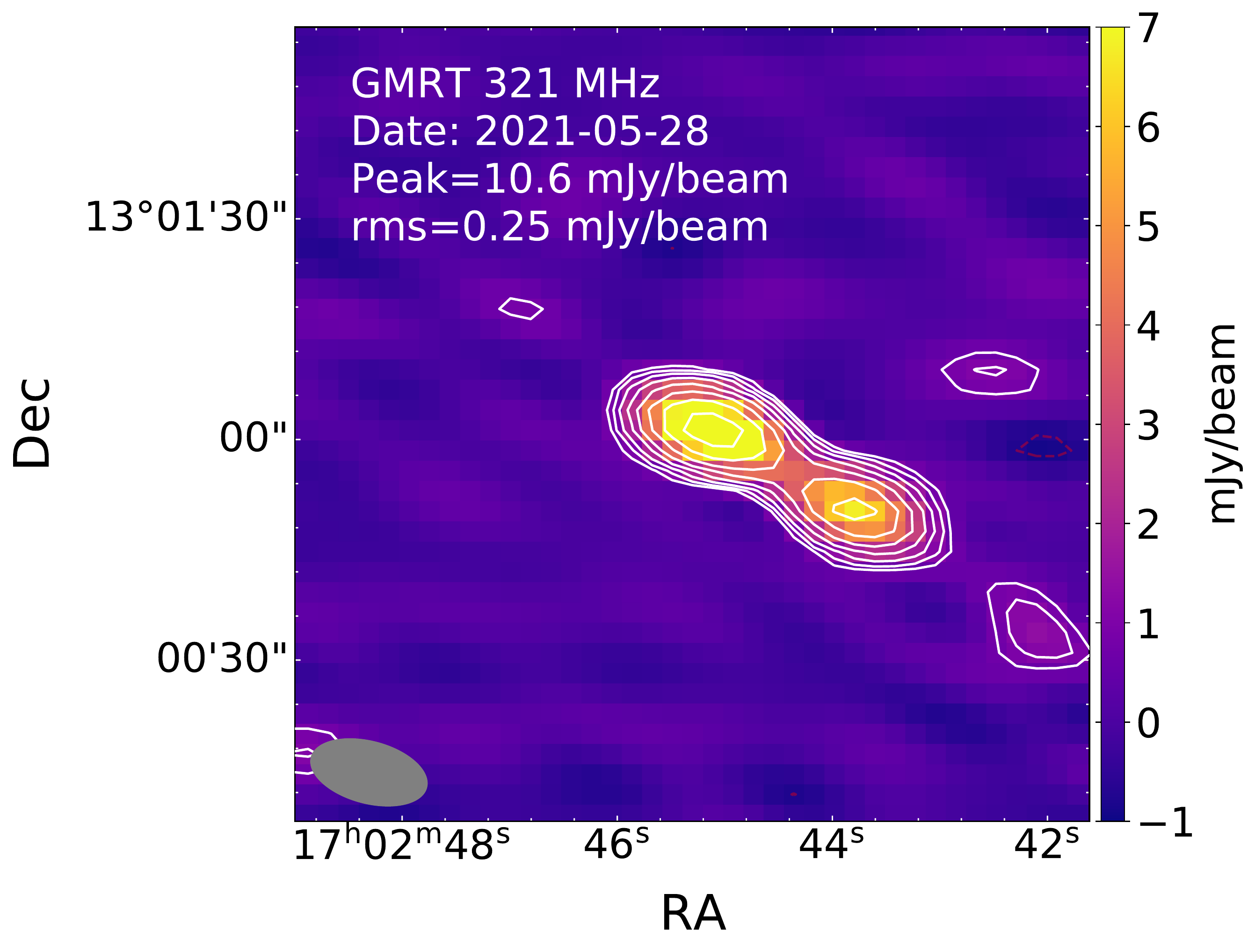}
    \caption{GMRT images of \tar. The observation date, frequency, peak intensity, and \textit{rms} noise of the images are labelled in each panel. The restoring beam is shown as an ellipse in the bottom-left corner. Contours start at three times \textit{rms} noise and increase in a step of 2.
    }
    \label{fig:GMRT}
\end{figure*}

The GMRT \citep{1991ASPC...19..376S} observations were carried out at 321, 607, and 1383 MHz on 2021 May 28, May 30, and June 7, respectively (proposal id: DDT179). A bandwidth of 32 MHz centred at the frequencies mentioned above was used in the observations. The GMRT Software Backend (GSB) was used as the backend correlator. The on-source times were $\approx 30$ mins, $\approx 29$ mins, and $\approx 21$ mins in the three observed bands, respectively.  3C~48 was used as the flux density calibrator in the 321 MHz and 1383 MHz observations. For the 607 MHz observation, the sky position of 3C~48 during the observation was not suitable for this observation, so we used 3C~468.1 as the flux density calibrator. We used the spectral index model parameters of 3C~468.1 to estimate its flux density at 607 MHz, which was interpolated by fitting the literature data with a fourth-order polynomial function \citep[][]{2017A&A...597A..96R}. Further, 1640+123 was used as the phase calibrator for the observations at 321 MHz and 1383 MHz, and 1737+063 was used as the calibrator at 607 MHz. The complex bandpass functions of each telescope were calibrated using the flux density calibrator 3C~48 or 3C~468.1. Except for a few edge channels, the data from other channels were combined and averaged over frequencies to improve the signal-to-noise ratio (SNR). After applying phase calibration solutions from the phase calibrator to the target \tar, we extracted the data of \tar\ for self-calibration and imaging. The final images are shown in Fig.~\ref{fig:GMRT}. The \textit{rms} noise in the images is 0.25 mJy beam$^{-1}$ at 321 MHz, 0.08  mJy beam$^{-1}$ at 607 MHz, and 0.13 mJy beam$^{-1}$ at 1383 MHz, respectively.

We fit Gaussian models to the source components in all radio images using the task IMFIT integrated into the Common Astronomy Software Applications (CASA, \citealt{2007ASPC..376..127M}). The model fit is conducted in the image domain which exports the peak flux density and right ascension (RA) and declination (Dec) positions, integrated flux density, and deconvolved size (Full Width at Half Maximum, FWHM) of the radio component.
The fitted parameters are listed in Table~\ref{tab:fitpars}. If the fitted component leads to an unresolved point source model, the upper limit to the component size is estimated using the component SNR and beam information of the image according to the relation given by \citet{2005astro.ph..3225L}. 
In addition to the model fitting error, we also considered the systematic error due to the visibility amplitude calibration, which typically accounts for a few per cent of the measured flux density. 

Figure~\ref{fig:GMRT} shows the 1.4 GHz, 610 MHz, and 321 MHz GMRT images. 
The unresolved NVSS source is clearly resolved into two components, with the northeast (NE) component being brighter and more compact than the southwest (SW) component. 
The NE component is close to the optical position of \tar, and the SW component is close to an infrared (IR) source WISEA J170243.84+130052.0 (J1702-SW hereafter). The separation between NE and SW is about 23.5\arcsec, corresponding to a projected size of $\sim$145.7 kpc at $z=5.5$ \footnote{ Using the cosmological parameters derived from a flat $\Lambda$ Cold Dark Matter ($\Lambda$CDM) model \citep{2011ApJS..192...18K} with $\Omega_\mathrm{m} = 0.27$, $\Omega_{\Lambda} = 0.73$, and $H_{0} = 70$~km s$^{-1}$Mpc$^{-1}$, 1 arcsec angular separation corresponds to a projected linear size of $\sim$6.2 kpc at $z=5.5$.}. A $4\sigma$ component is marginally detected at the \tar\ position in the TGSS image.
In the 321-MHz image, the two components are connected. However, in the 1.4 GHz and 610 MHz images, the connection between the two components is missing. 
It is not possible to distinguish from the images alone whether NE and SW belong to one source or two physically unrelated sources, nor is it possible to determine whether it is a double-lobed Fanaroff-Riley type 2 (FR II) source or a one-sided core-jet source. We discuss the radio structure in detail in Section~\ref{sec:discussion}.

\subsection{ASKAP} 

\begin{figure}
    \centering
    \includegraphics[width=0.4\textwidth]{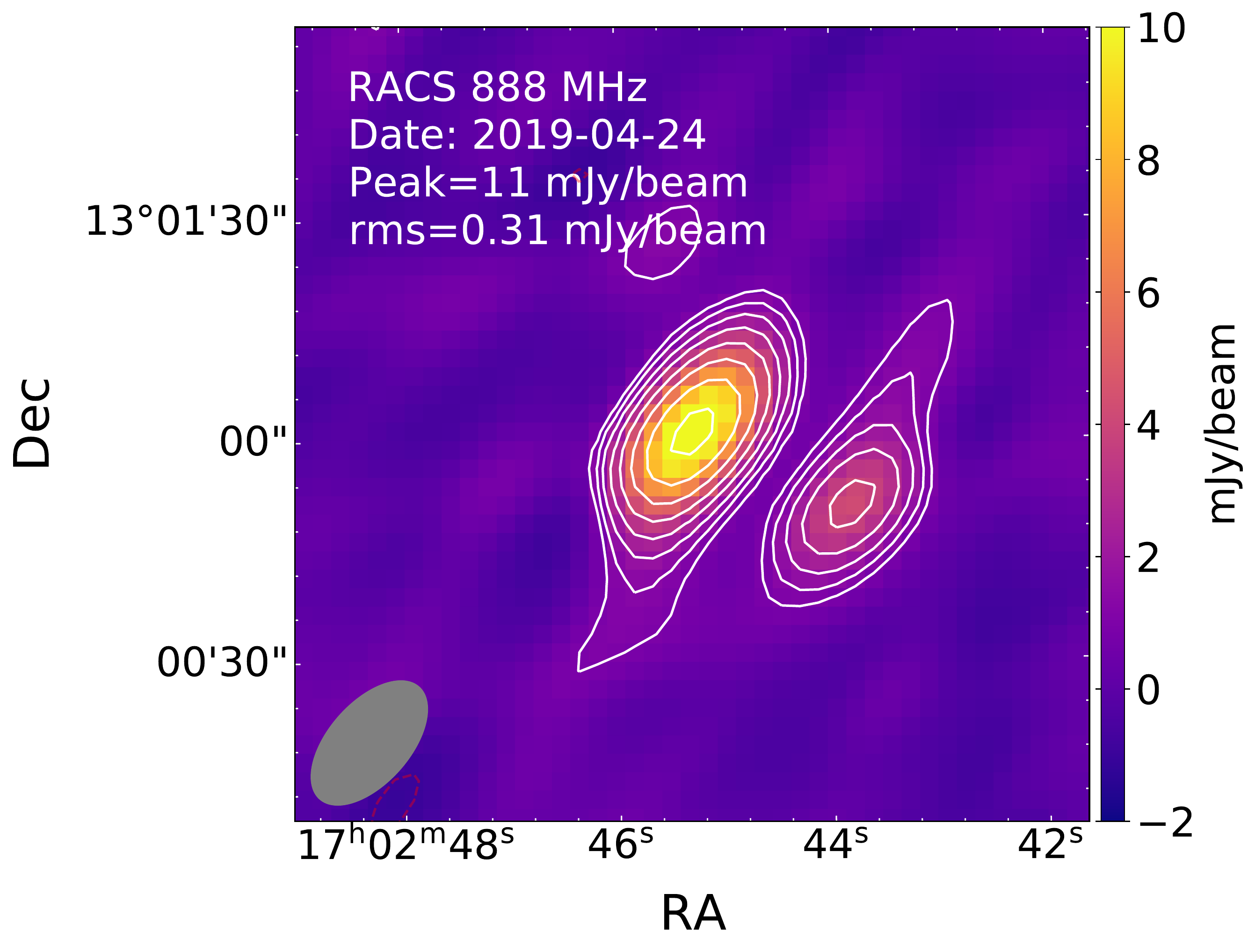}
    \includegraphics[width=0.4\textwidth]{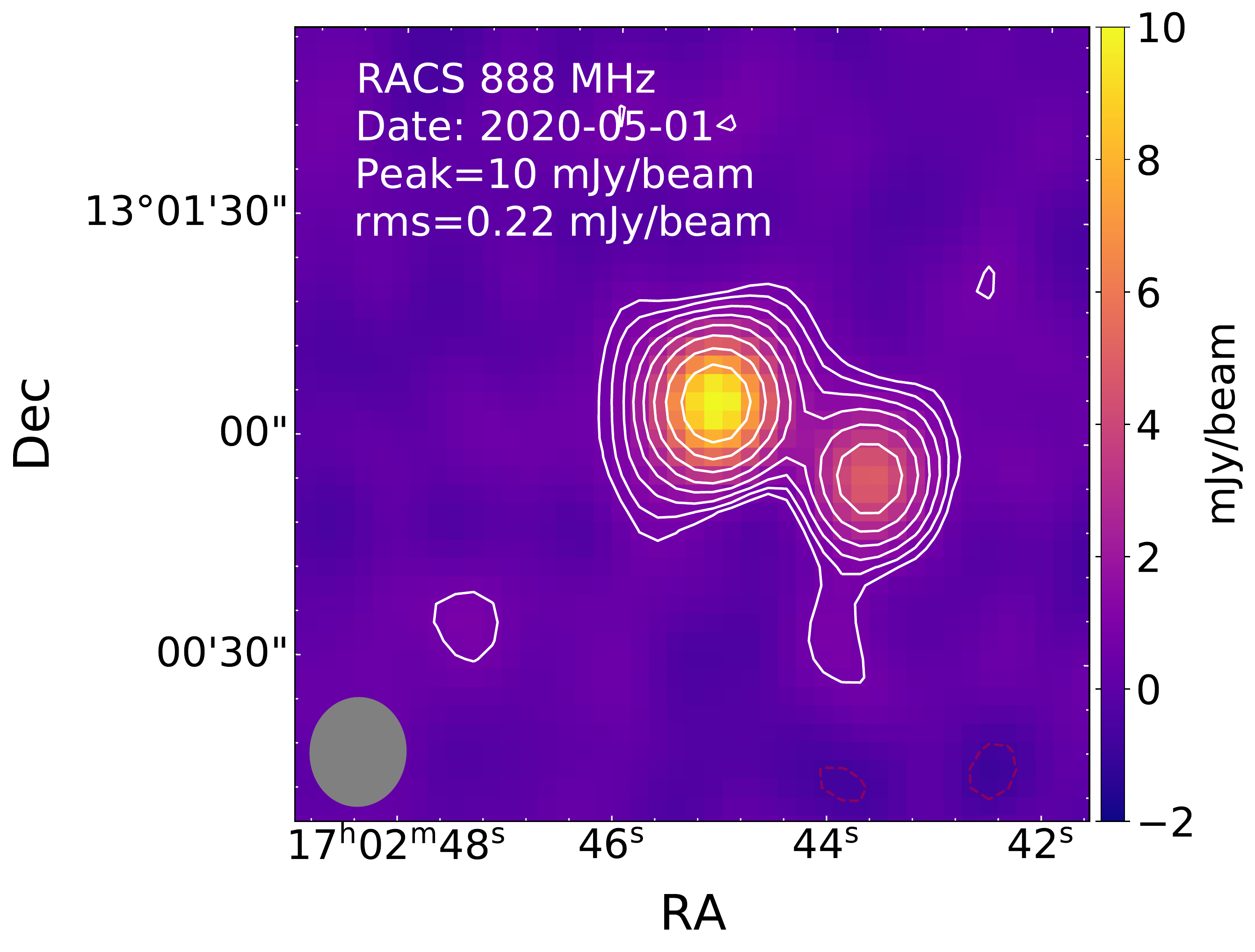}
    \caption{ASKAP images of \tar\ at 888 MHz.  The observation date, frequency, peak intensity, and \textit{rms} noise of the images are labelled in each panel. The restoring beam ($20.3\arcsec\times10.9\arcsec$ and $14.7\arcsec\times13.0\arcsec$ in the top and bottom panels, respectively) is shown as an ellipse in the bottom-left corner. Contours start at three times \textit{rms} noise and increase in a step of 2. }
    \label{fig:ASKAP}
\end{figure}

The ASKAP \citep{2021PASA...38....9H} is a precursor telescope of the SKA-Mid. 
The Rapid ASKAP Continuum Survey (RACS) is the first shallow large-area survey conducted with all 36 antenna before the deep surveys are carried out with the full 36-antenna ASKAP \citep{2020PASA...37...48M}.
The RACS observations began on 2019 April 21 and finished on 2020 June 21. 
RACS covers the sky south of declination +41$\degr$  over a 288-MHz band centred at 887.5 MHz.  
In most sky zones, the median \textit{rms} noise is $\sigma_{\rm med} = 0.25 \, {\rm mJy} \, {\rm beam}^{-1}$.
We found images of \tar\ in the RASC archive\footnote{CASDA Observation Search: \url{https://data.csiro.au/domain/casdaObservation}.} on 2019 April 24 and 2020 May 1 (Fig.~\ref{fig:ASKAP}). 
The peak flux density is 11 mJy beam$^{-1}$ and 10 mJy beam$^{-1}$ in the 2019 and 2020 images, respectively.
The resolution of the RACS images is $\sim$15\arcsec, between the resolutions of the 1.4-GHz and 0.61-GHz GMRT images; all these images consistently show two identical radio components associated with \tar\ and J1702-SW, respectively. 

\subsection{VLA}

We downloaded the image of \tar\ from the NVSS Postage Stamp Server  \footnote{\url{https://www.cv.nrao.edu/nvss/postage.shtml}}. The NVSS covers almost all of the sky north of ${\rm DEC}=-40\degr$ at a central frequency of 1.4 GHz \citep{1998AJ....115.1693C}.  The angular resolution of NVSS is $45\arcsec$, and the \textit{rms} noise in the total intensity images is $\sigma \approx 0.45$ mJy beam$^{-1}$.
The radio counterpart of \tar\ is detected at RA=17h02m45.18s, Dec=+13d01m01.0s, consistent with its optical position within $3\sigma_{\rm p}$. 
The peak and integrated flux density of \tar\ is 24.4 mJy beam$^{-1}$ and $27.8\pm0.4$ mJy, respectively.

\begin{figure*}
    \centering
    \includegraphics[width=0.3\textwidth]{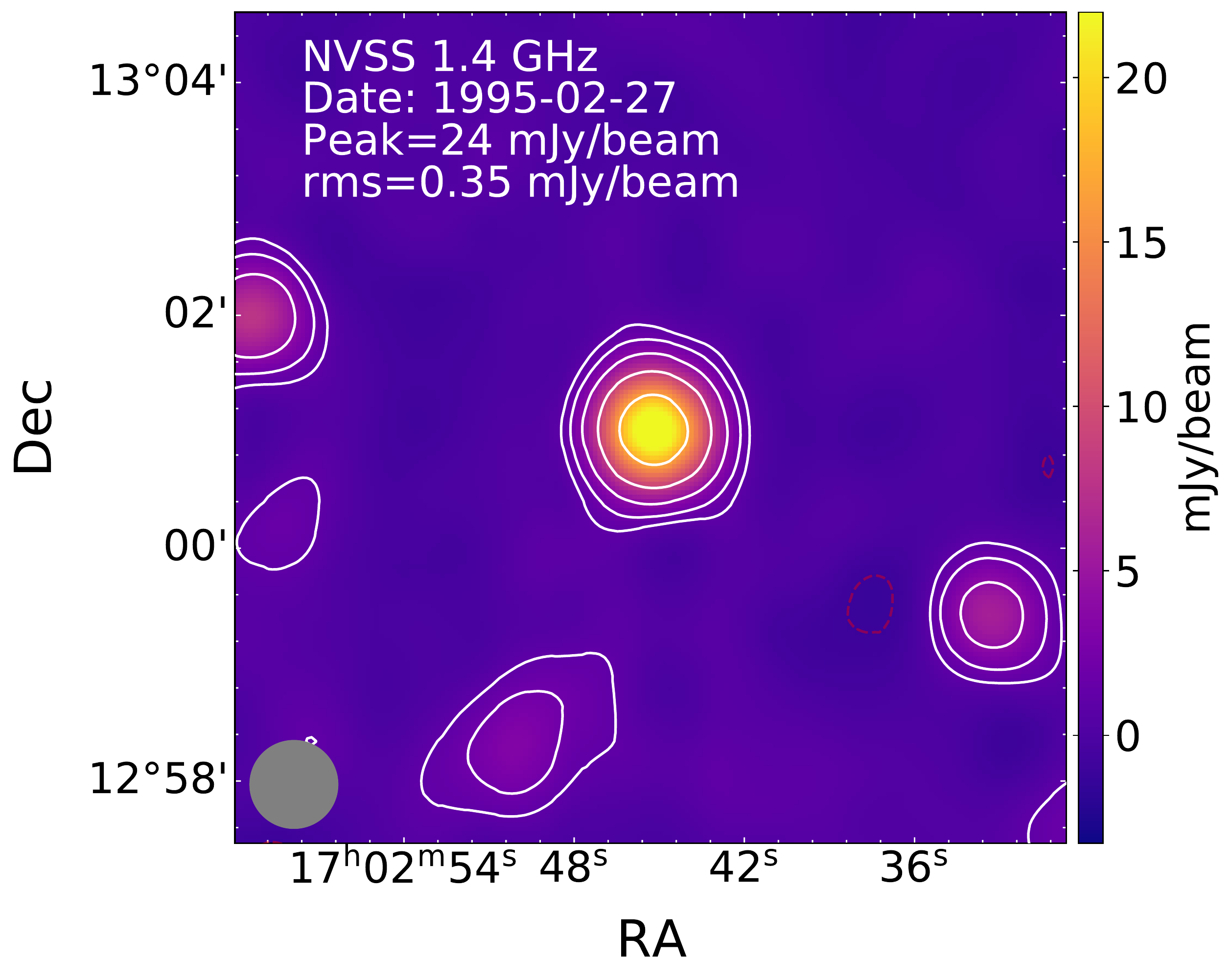}     \includegraphics[width=0.33\textwidth]{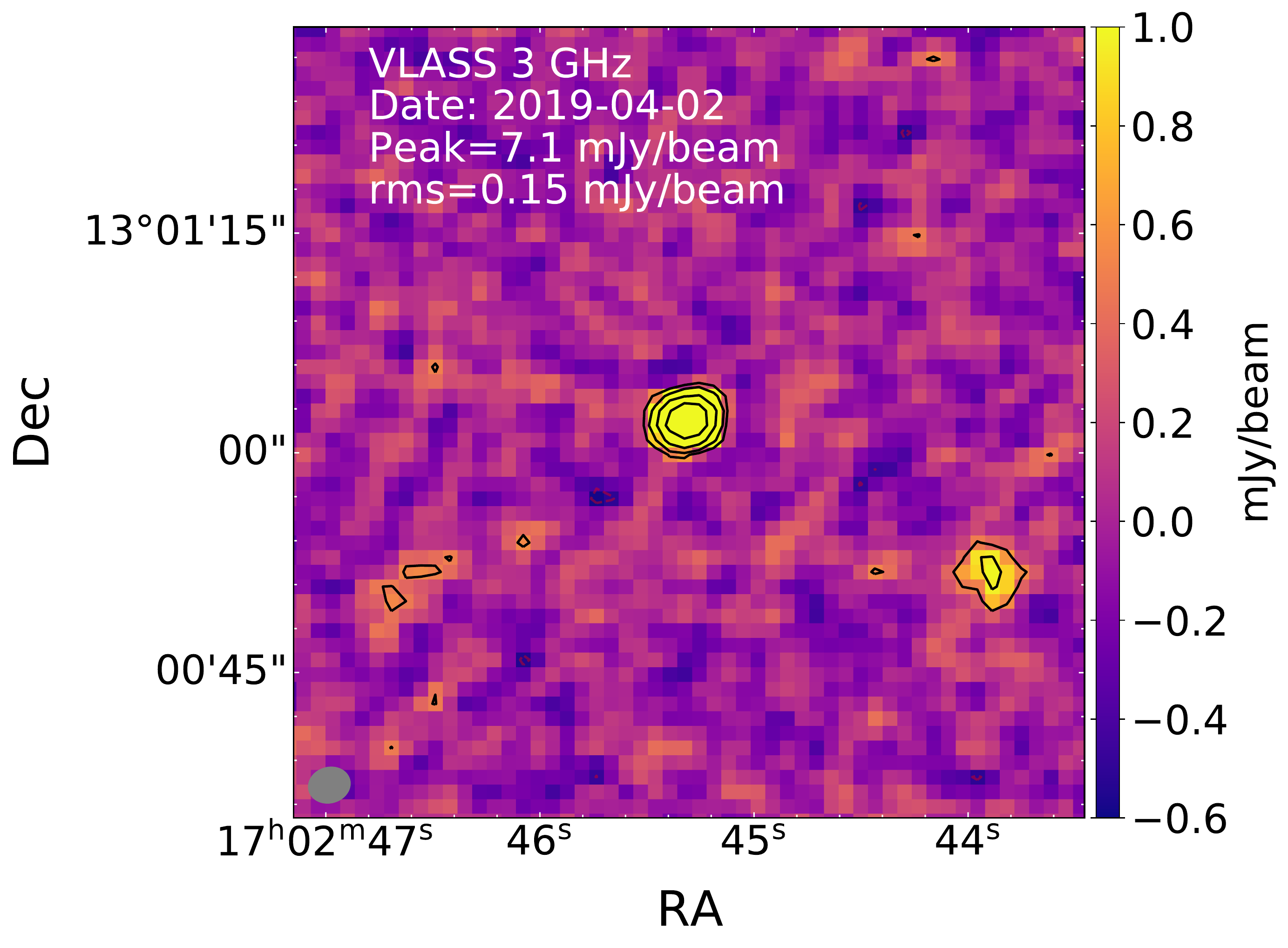}
    \includegraphics[width=0.33\textwidth]{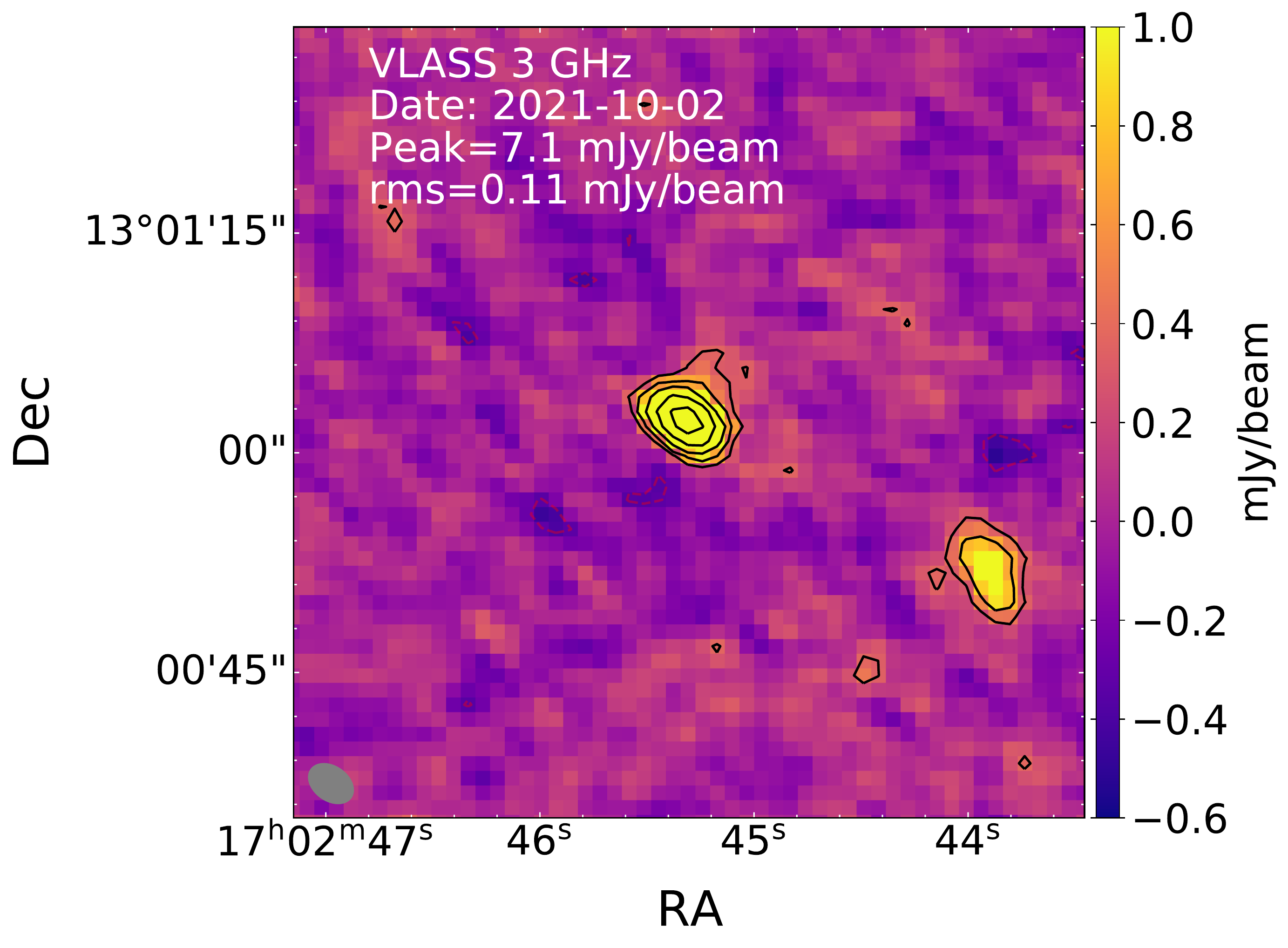}    
    \caption{VLA images of \tar. \textit{Left}: NVSS 1.4 GHz, contours represent $3\sigma \times (1,2,4,8,16)$; \textit{Middle}: VLASS 3 GHz observed in 2019, contours represent 0.45 mJy beam$^{-1} \times (1,2,4,8)$; \textit{Right}: VLASS 3 GHz in 2021, contours represent 0.33 mJy beam$^{-1} \times (1,2,4,8,16)$. }
    \label{fig:VLA}
\end{figure*}

The Very Large Array Sky Survey \citep[VLASS, ][]{2020PASP..132c5001L} is another large-area continuum survey made by the VLA after NVSS with an increased observing frequency of 3 GHz and a higher angular resolution of 2.5\arcsec\ (a factor of 18 higher than that of NVSS). VLASS covers the same sky area as NVSS.
The sensitivity of the stacked images is $\sim 70 \, \mu$Jy. 
In the VLASS images (Fig.~\ref{fig:VLA}), the NE component is clearly detected above $50\sigma$.
J1702-SW is  weaker but still detected with 6.6$\sigma$--10.0$\sigma$. The morphology of NE is compact, and SW shows a resolved morphology. 
No significant feature is detected between the NE and SW components.

\subsection{RATAN-600} 

The RATAN-600 is a meridian radio telescope with a ring-shaped variable profile antenna \citep{1979S&T....57..324K,1993IAPM...35....7P,2020gbar.conf...32S} allowing to obtain radio spectra at the frequencies of 2.3, 4.7, 8.2, 11.2, and 22.3 GHz quasi-simultaneously within 3--5 min. The angular resolution depends on the declination of the source. For J1702+1301, the angular resolution in right ascension is 48\arcsec, 26\arcsec\ and 20\arcsec\ at 4.7, 8.2 and 11.2 GHz, respectively; the angular resolutions in declination are $\sim$4 times worse than in RA. Observations of J1702+1301 were made in 15 epochs from 2020 December to 2022 October, with 3–-31 measurements per observing epoch to improve the signal-to-noise ratio. The observations were processed using an automated data reduction system \citep{2016AstBu..71..496U} based on the Flexible Astronomical Data Processing System (FADPS) standard data reduction software \citep{1997ASPC..125...46V}. The observational statistics and measured flux densities for J1702+1301 are presented in Table~\ref{tab:ratan}.

\begin{figure*}
    \includegraphics[width=0.68\textwidth]{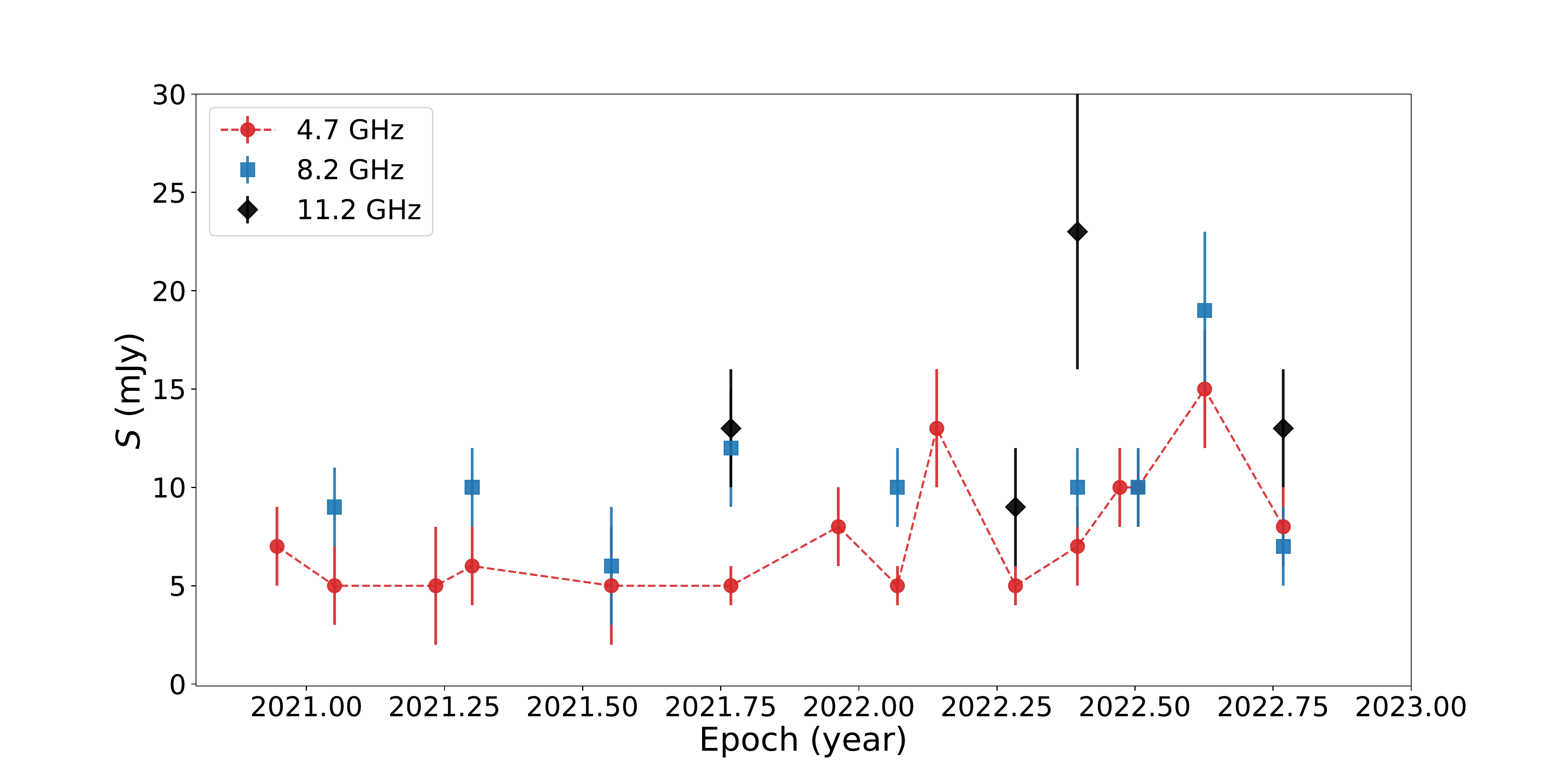}  
    \includegraphics[width=0.3\textwidth]{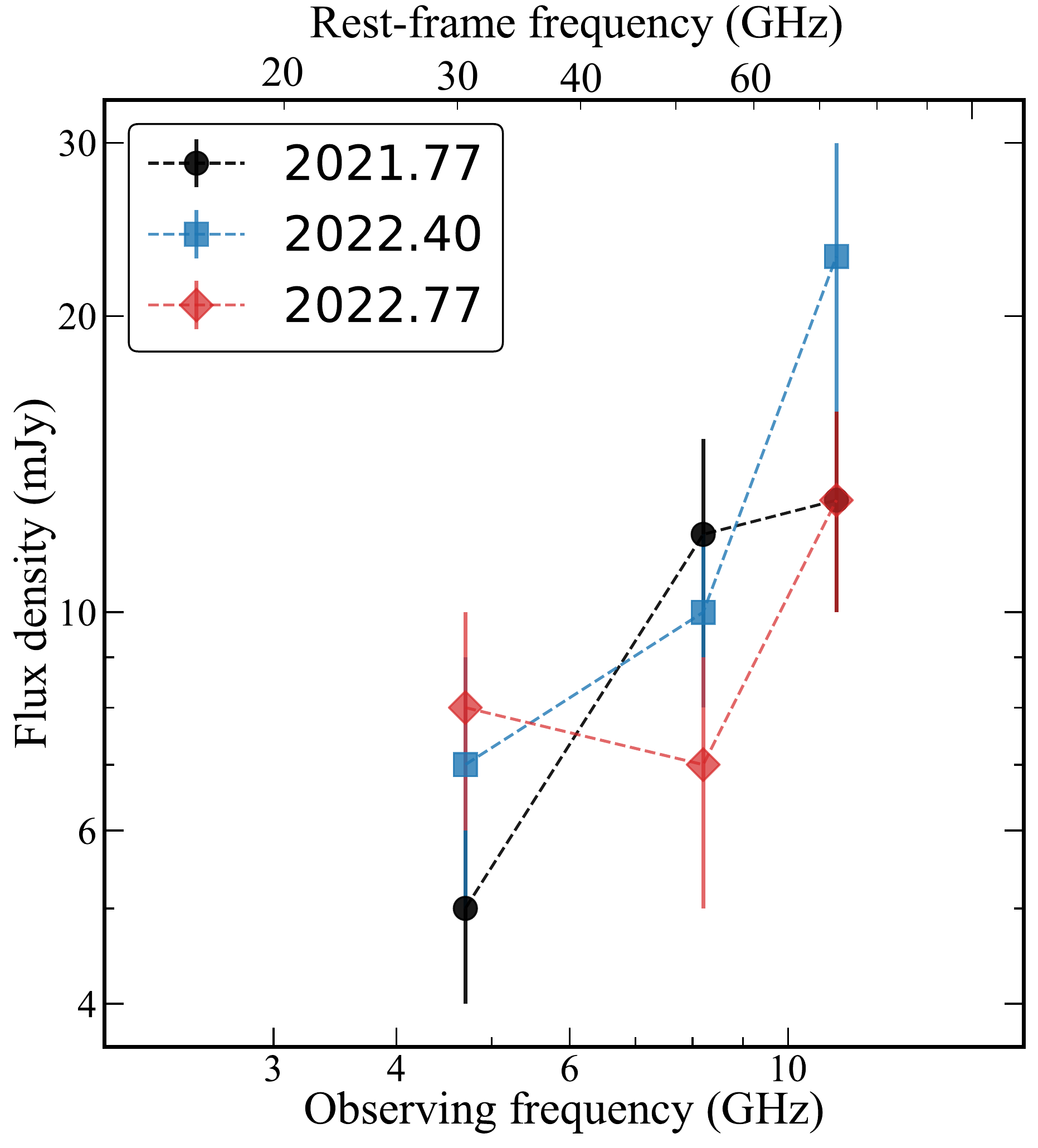}
 \caption{RATAN-600 light curves of J1702+1301 from 2020 December to 2022 October (left) and on selected dates (right).}
    \label{fig:ratan}
\end{figure*}

We know from the analysis in Section~\ref{sec:discussion} that J1702-SW is a steep-spectrum radio source that contributes a small fraction to the total flux density at 5 GHz and above.
Therefore, we can assume that most of the RATAN-600 flux densities at $\nu \gtrsim\ 5$GHz come from J1702+1301.
The light curves are shown in Figure~\ref{fig:ratan}.  
For some epochs without sufficiently significant signals ($\rm SNR \leq 3$), the data are not included.  
This quasar demonstrates moderate variability (30--40 per cent) at 4.7, 8.2 and 11.2 GHz over a monitoring period of almost two years, with the fractional variability values: F$_{\rm var}$=0.41$\pm$0.06 at 4.7 GHz, F$_{\rm var}$=0.35$\pm$0.03 at 8.2 GHz. 
F$_{\rm var}$ is calculated by the relation from \cite{2003MNRAS.345.1271V}. 
At 4.7 GHz, the highest variability amplitude is approximately 2.5 times the average value in the quiescent state ($\sim 6$ mJy), with the bursts lasting 1--4 months (corresponding to 4--19 days in the source's rest frame). The average flux density at 8.2 GHz in the quiescent state is about 10 mJy, and the strongest flare is 1.9 times higher. The data sampling at 11.2 GHz is sparse and does not allow an accurate assessment of the quiescent-state flux density. The maximum flux density at 11.2 GHz is 23 mJy. The three epochs with pronounced variability are highlighted in the right pane of Fig.~\ref{fig:ratan}.

\section{Discussion}\label{sec:discussion}

\subsection{Radio morphology}

\begin{figure*}
    \centering
    \includegraphics[width=0.35\textwidth]{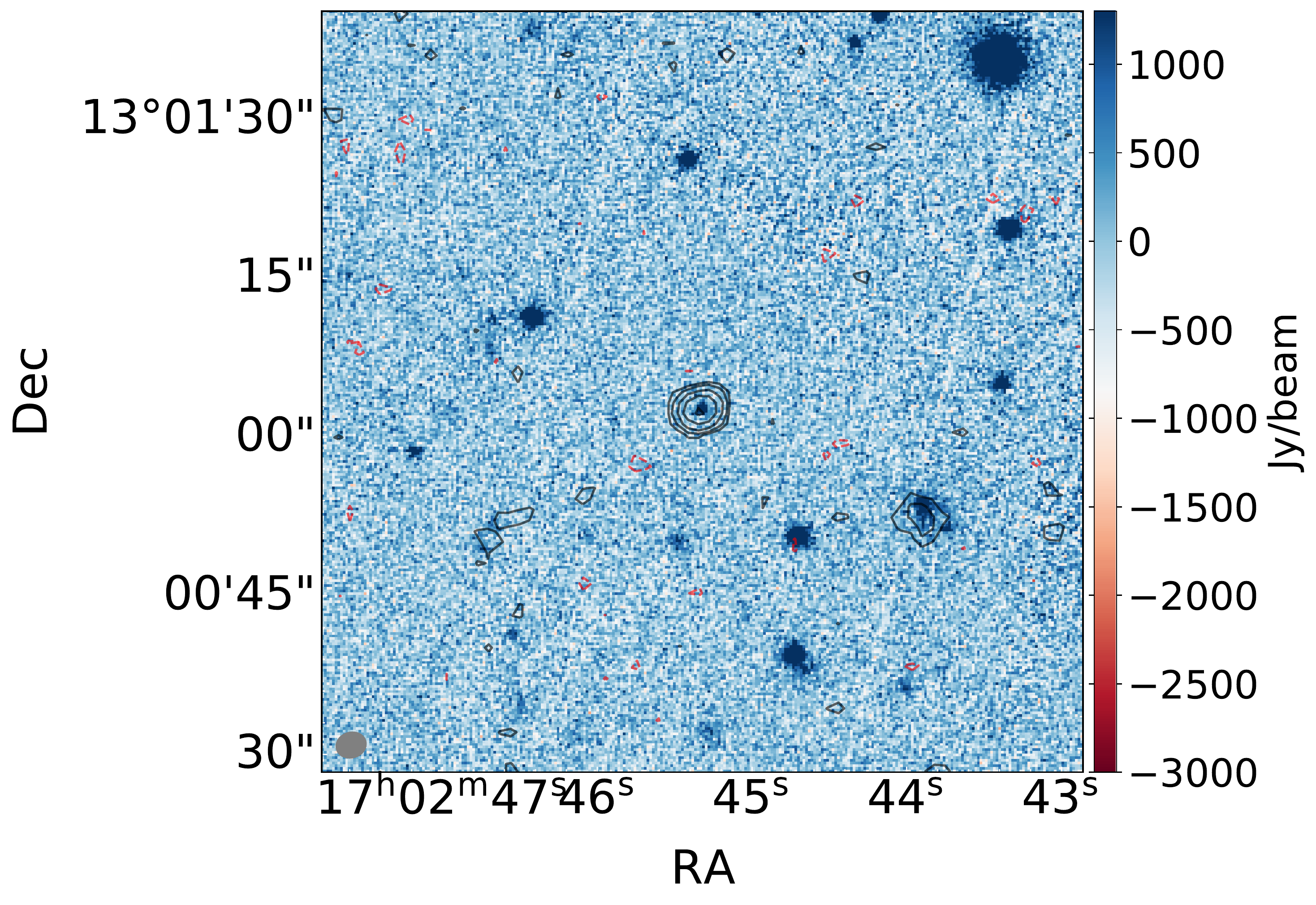}
    \includegraphics[width=0.33\textwidth]{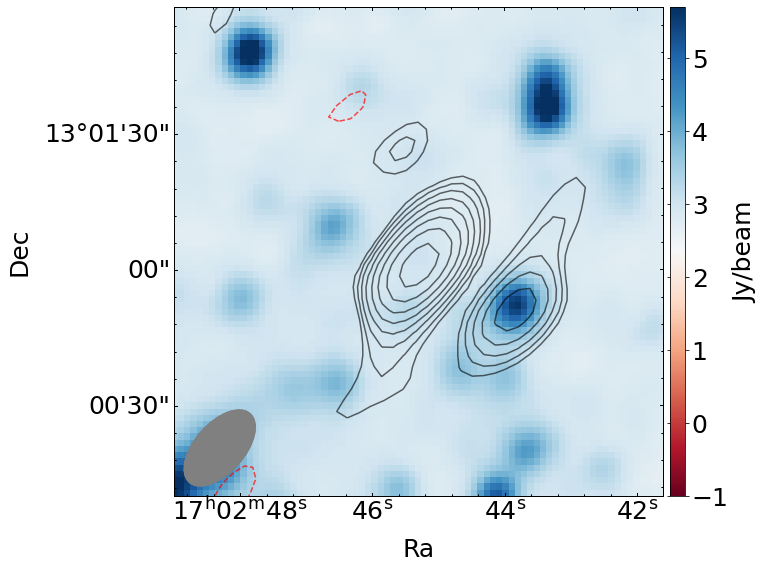}
    \includegraphics[width=0.24\textwidth]{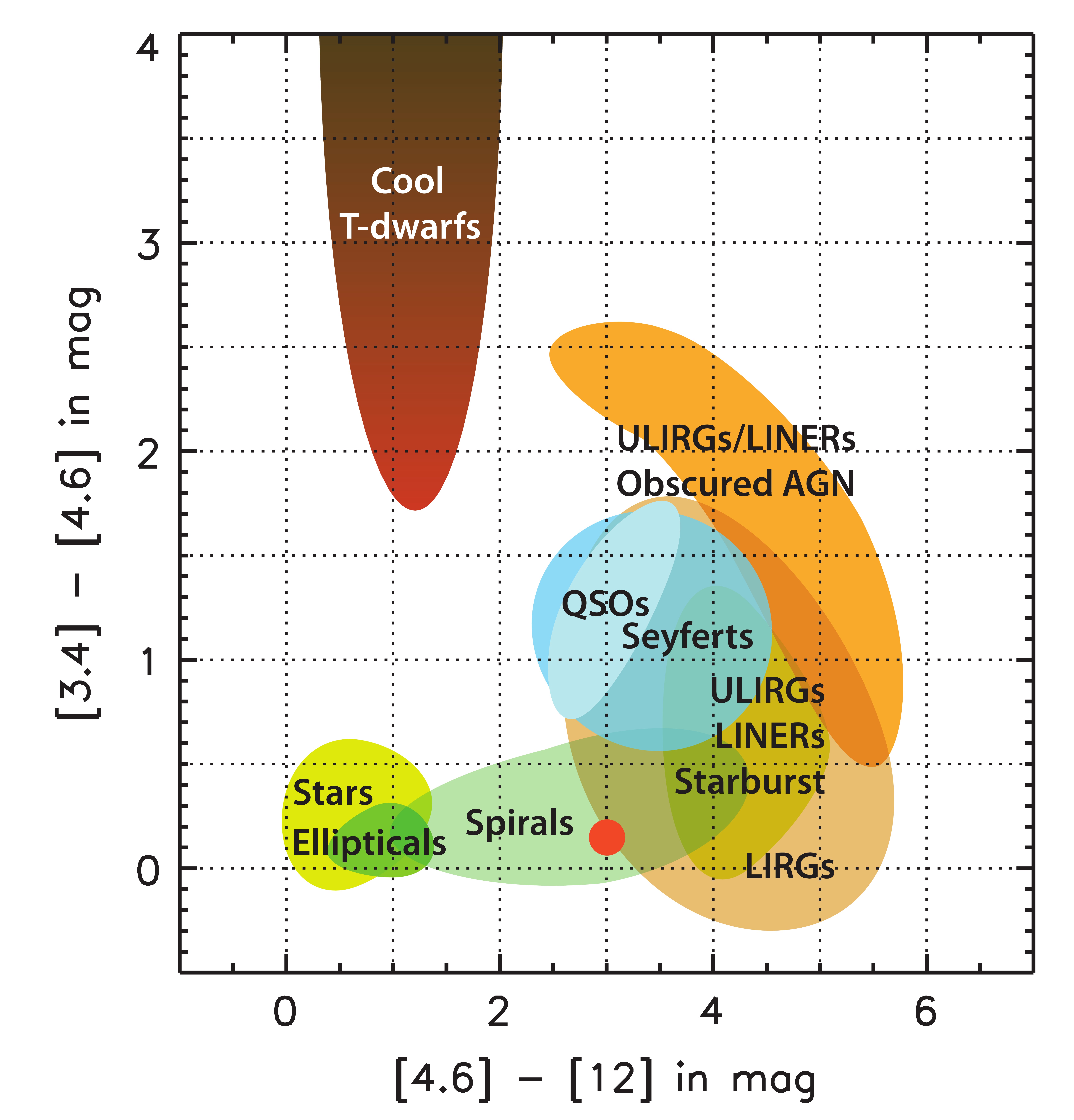}
    \caption{Optical Pan-STARRS1 images (left panel) at \textit{i}-band (colour scale) overlaid with the VLASS image in epoch 2019, WISE 3.4 $\mu$m image of  $1\arcmin \times 1\arcmin$ overlaid with the ASKAP contour image in epoch 2019 (middle panel) and IR colour-colour map (right panel). In the right panel, \tar\ is denoted by a solid red-coloured circle.
    }
    \label{fig:J1702-SW}
\end{figure*}

Details of the observations are given in Table \ref{tab:obs}, and Figures~\ref{fig:GLEAMX}--\ref{fig:VLA} show the radio structure of \tar\ and its surrounding region at different frequencies and with different resolutions. In low-resolution NVSS and MWA images, this region shows only an unresolved source. But it is resolved into two components in ASKAP, GMRT and VLASS images: \tar\ (the NE component) and J1702-SW (23.5\arcsec\ to the southwest). 

\textit{Could NE and SW be radio components of the same AGN?} 
The greatest difficulty in this scenario lies in interpreting the inferred ultra-long jet of \tar. According to the definition of a quasar in the AGN unified scheme, a core-dominated quasar has a viewing angle of less than 10\degr\ \citep{1993ApJ...407...65G}. If NE is the core and SW is the lobe, the separation between NE and SW is 145 kpc, indicating that the intrinsic jet length will exceed 840 kpc and making \tar\ the only high-redshift $z>5$ quasar observed with an Mpc-scale jet. This jet length is very rare in the high-redshift Universe and is three orders of magnitude larger than that previously found in the $z = 5.84$ quasar PSO J352.4034-15.3373 \citep{2018ApJ...861...86M}. The strength of the inverse Compton scattering of the cosmic microwave background (CMB) photons increases with $(1+z)^4$, rendering the survival of extended jets in the high-$z$ Universe a great challenge \citep{2015MNRAS.452.3457G}. Moreover, the dense interstellar medium of the host galaxies of the high-$z$ AGN forms a substantial obstacle to the jet growth \citep{2020NatCo..11..143A,2022MNRAS.511.4572A}, limiting a significant fraction (even majority) of the jets to a few kpc \citep{2004NewAR..48.1157F}.
Therefore, it is unlikely that NE and SW belong to one source. 

\textit{Are NE and SW  in a dual AGN system?} We show the radio (VLASS) and optical (PAN-STARRS1) overlaid images in the left panel of Fig.~\ref{fig:J1702-SW}. 
The optical morphology of SW looks diffuse and inconsistent with the high-redshift quasars observed so far. 
We then estimated the photometric redshift of J1702-SW to be 0.677 based on the Sloan Digital Sky Survey \citep[SDSS, ][]{2022ApJS..259...35A} and Wide-field Infrared Survey Explorer (WISE) photometry data \citep{2014yCat.2328....0C} using a public software program of Easy and Accurate Redshifts from Yale (EAZY, \citealt{2008ApJ...686.1503B}).  We used a set of nine templates from \cite{2008ApJ...686.1503B} that combine the Simple Stellar Population Models \citep{2009ApJ...699..486C,2005MNRAS.362..799M} with varying dust attenuation and ages along with emission line models. The photometric redshift obtained using the DECaLS data \citep{2019AJ....157..168D} for the SED fitting is slightly lower. However, the southwest source displays multiple structures in the DECaLS $z$-band image, which causes a larger uncertainty in the estimation of the photometric redshift.
Therefore, the photometric redshift results suggest that J1702-SW is a foreground galaxy.

We calculated WISE $W1-W2$ and $W2-W3$ of J1702-SW. We found that it is located in the transition region between spiral and starburst galaxies in the IR colour-colour diagram (Fig. \ref{fig:J1702-SW}), indicating that it is an active galaxy.
With this photometric redshift, we calculate its luminosity at rest-frame 1.4 GHz to be $9.7\times 10^{40} {\rm erg \ s^{-1} }$, and the corresponding star formation rate (SFR) is 4400 $\rm M_{\odot}$ yr$^{-1}$. The inferred SFR exceeds that of the most active starburst galaxy Arp~220 \citep{1998ApJ...493L..17S,2008MNRAS.384..875R} by a factor of 20--40, suggesting that the radio flux of J1702-SW could come from the combined contribution of star formation and AGN.

\subsection{Is J1702+1301 a blazar?}

\begin{figure}
    \centering
    \includegraphics[width=0.5\textwidth]{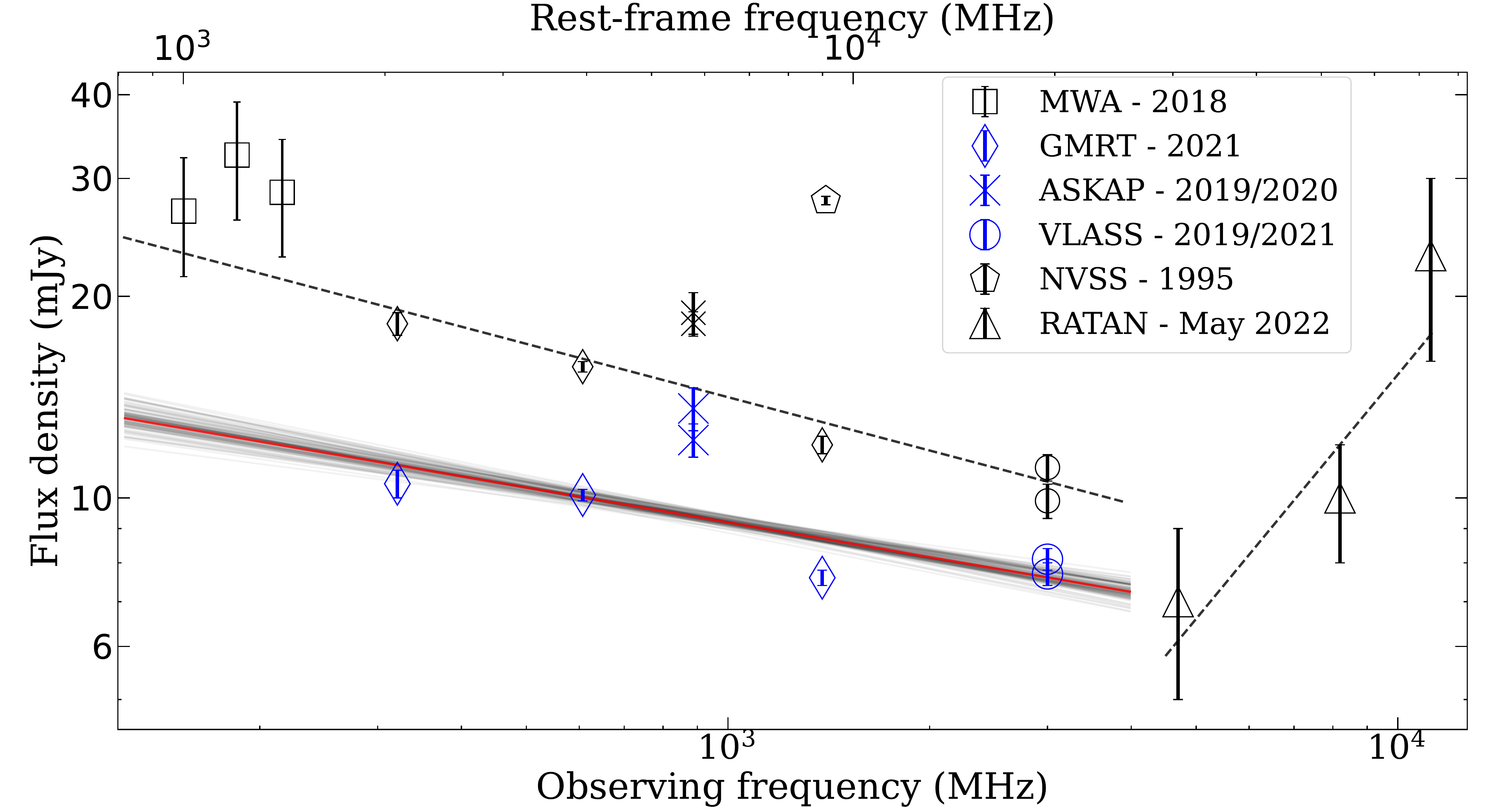}
    \caption{Non-simultaneous radio spectrum of \obj. 
    The data points are from MWA GLEAM-X, GMRT, ASKAP RACS, VLA NVSS and VLASS, and RATAN-600 observations.
    The blue-coloured symbols represent the \obj\ (NE component in GMRT and ASKAP images).  The spectral indices are obtained by fitting the spectra with power-law functions: $\alpha_{\rm NE}=-0.17\pm0.05$. The black-coloured symbols represent the combined flux densities of \tar\ and J1702-SW.}.
    \label{fig:1702sed}
\end{figure}

Figure~\ref{fig:1702sed} shows the radio spectrum of \tar. 
The overall radio spectrum is plotted as a comparison. 
The overall spectrum exhibits a clear truncation around 5 GHz: $<5$ GHz, the spectrum is flat; $>5$ GHz, it shows a rising spectral shape.
J1702-SW has a steeper spectrum with a spectral index of $\alpha= -0.47\pm 0.12$,
and its flux density decreases rapidly above 5 GHz: no higher than 2.3 mJy at 5 GHz and less than 1.8 mJy at 8 GHz. Therefore, the RATAN-600 measurements can approximately reveal the high-frequency radio spectrum of \tar. 

The spectral index of \tar\ is obtained by fitting the radio spectrum from 300 MHz to 3 GHz using a power-law function as $\alpha_{\rm J1702+1301} = -0.17 \pm 0.05$. 
The spectral fit is based on non-simultaneous observations distributed from April 2019 to October 2021. It basically reflects the flat spectrum characteristics of \tar\ albeit with the effect of optical variability.  The uncertainty caused by variability is already included in the error of the spectral index.

The variability magnitude of \tar\ in low-frequency  ($\nu \leq 3$ GHz) bands is insignificant on $\sim$2-year time period. For example, the flux density of the NE component (J1702+1301) at 3 GHz increased from 7.7 mJy in April 2019 to 8.1 mJy in October 2021, and changed only 10 per cent at 888 MHz on a one-year timescale. However, its long-term  variability at lower radio frequencies and the strong short-term variability at higher radio frequencies and high energy bands are worthy of attention.
Comparing the total flux density obtained by the GMRT at 1.4 GHz in 2021 with that obtained by the VLA NVSS in 1995, we found a 2.3-fold decrease in flux density. Due to the steep spectrum feature of J1702-SW (dominated by optically thin jet and star formation), we have reason to believe that the 1.4-GHz variability over a time span of $\sim$26 years (4 years in the source's rest frame) is mainly from \tar. At 4.7 GHz (corresponding to $\nu_{\rm rest} \sim$30 GHz) and above, the RATAN-600 light curves in Fig.~\ref{fig:ratan} demonstrate prominent variability on time scales of 4--19 days, with variation scales of more than a factor of 2.
Moreover, \citet{2021AstL...47..123K} reported a 2-fold variability in X-rays of \tar\ over a six-month period ($\Delta t_{\rm rest} \lesssim$ a month). 
\tar\ exhibits significant variability at different time scales in the X-ray and radio bands, and the time scales of the short-term variability obtained from the X-ray and high radio frequencies are consistent with typical blazars. These variability properties support J1702+1301 being a blazar.

\citet{2021AstL...47..123K} used the NVSS flux density at 1.4 GHz and an assumed flat spectral index $\alpha = 0$ to estimate the 5-GHz flux density in the source's rest frame, $S_{\rm 5GHz, rest} = 9.6 $ mJy. It is known from our analysis above that the NVSS flux density contains contributions from both NE and SW components. Therefore, it is an overestimate for the flux density of \tar. Moreover, our Figure~\ref{fig:1702sed} shows a radio spectrum of \tar\ with $\alpha_{\rm NE} = -0.17$ between 0.3 and 3 GHz, which allows us to obtain a more accurate estimate of its flux density $S_{\rm 5GHz,rest} = 9.6$~mJy.

The rest-frame wavelength $4400\angstrom$ for the source at $z=5.5$ corresponds to the observing wavelength of 2.8 $\mu m$, close to the wavelength of the \textit{Wide-Field Infrared Survey} ($WISE$, \citealt{Wright2010}) $W1$ band. However,  \textit{WISE} did not detect \tar\ in its \textit{W1} band. If we use the detection limit of \textit{W1} magnitude (Vega) of 18.84 to constrain the rest-frame optical luminosity $S_{\rm 4400 \angstrom}$, we obtain a lower limit of the radio loudness parameter $R > 1100$.

The intrinsic radio luminosity of \tar\ at $\nu_{\rm 5GH, rest}$ GHz derived from the 2020 ASKAP observation is ($3.8 \pm 0.2)\times 10^{43}$ erg s$^{-1}$. Its historical luminosity in the 1995 NVSS image is even higher (see the discussion of variability above). 
The high radio luminosity makes it outstanding in $z>5$ AGNs \citep{2015ApJ...804..118B,2020A&A...635L...7B,2021MNRAS.508.2798S}.
We note that several $z > 5$ quasars with $L_{\rm 5GHz}> 2 \times 10^{43}$~erg~s$^{-1}$ have all been identified as blazars, e.g., B2 1023+25 \citep[z=5.3,][]{2012MNRAS.426L..91S,2015MNRAS.446.2921F},
Q0906+6930 \citep[$z=5.47$, ][]{2020NatCo..11..143A}, PSO J047.4478+27.2992 \citep[$z=6.1$, ][]{2020A&A...635L...7B,2020A&A...643L..12S}. 
The observational characteristics of \tar, such as significant and rapid variability, high radio and X-ray luminosity, and high radio loudness, are all consistent with its blazar identification. Future observational evidence from X-rays and radio will strengthen its identification as a blazar.

\subsection{Comparison of J1702+1301 and J1429+5447}

\tar\ is one of the two brightest sources discovered in the eROSITA survey and has the highest X-ray luminosity among the $z>5$ quasars. The other eROSITA-detected $z>5$ quasar is CFHQS J142952+544717 ($z=6.18$, J1429+5447 in short, \citealt{2020MNRAS.497.1842M}). The X-ray luminosity, bolometric luminosity, and black hole mass of \tar\ and J1429+5447 are very similar. They also share some common radio properties: they have similar radio luminosity; J1429+5447 shows a compact structure on arcsec scales and remains unresolved even on the 100-pc scale \citep{2011A&A...531L...5F}, and \tar\ in the VLASS images remains compact on sub-arcsec scales.
Whereas, they have some different observing characteristics: J1429+5447 has a steep radio spectrum, while the radio spectrum of \tar\ is flat;
J1429+5447 has no strong variability, while \tar\ shows significant radio variability on weekly timescales and X-ray variability on monthly timescales;
the radio loudness of \tar\ is about one order of magnitude higher than that of J1429+5447. 
These differences between \tar\ and J1429+5447 may be related to the discrepancy in the relativistic beaming effects of their jets.

In addition, the optical-X-ray spectra of \tar\ and J1429+5447 are flatter than that of radio-loud quasars (i.e., $\alpha_{\rm ox}$ is lower), suggesting an additional contribution of X-rays from the jet \citep{2011ApJ...726...20M,2019MNRAS.489.2732I}. This jet-related X-ray excess is found to become more pronounced in high-redshift objects \citep{2013ApJ...763..109W,2019MNRAS.482.2016Z}, which is associated with the enhanced inverse Compton scattering with redshift. Studying \tar\ in multiple bands is essential for understanding the jet physics in extremely-luminous high-redshift X-ray sources.

\section{Conclusion}
SRGE J170245.3+130104 is a high-redshift quasar discovered in the X-ray survey and confirmed by the follow-up optical spectroscopic observations \citep{2021AstL...47..123K}. It is a typical representative of extremely luminous active galactic nuclei in the early Universe and is the tip of the iceberg of high-redshift quasars. Based on the newly observed radio data and archive data, we performed a comprehensive study of the radio properties of \tar. In the high-resolution images, the radio counterpart previously detected in the NVSS image is resolved into two components separated by about 23.5\arcsec, where the northeast component corresponds to the optical and X-ray positions of \tar. 
The southwest component is associated with a foreground infrared galaxy with an insignificant contribution to the high-frequency spectrum of \tar. 
Our high-resolution radio data allow an accurate determination of the radio spectrum of J1702+1301.
J1702+1301 shows a flat radio spectrum and can be classified as a flat-spectrum radio-loud quasar. 
Significant variability has been observed in \tar\ in both X-ray and radio bands. 
We recalculated the radio loudness of J1702+1301 as $R> 1100$. It has a remarkable radio luminosity among high-redshift quasars, and the high-frequency ($\nu_{\rm rest}>30$ GHz) radio emission has rapid and large variability, indicating active jet activity.
The observed radio properties of \tar\ provide support for it being a blazar.
High-resolution Very long baseline interferometry (VLBI) polarimetric observations of \tar\ are critical for determining the beaming characteristics and geometry of the jet as well as the interactions of the jet with the interstellar medium.
An in-depth study of the jet properties of J1702+1301 will not only give a clear picture of the radiative properties of J1702+1301 itself but also help to understand the growth mode of black holes in the early Universe.

\begin{table*}
    \centering
    \caption{Model fitting parameters.}
    \begin{tabular}{cccccccc}\hline\hline
    Obs-date      &Telescope         &$\nu$    &Comp.        &RA                 &Dec               &$S_{\rm int}$      &Deconvolved Size                                        \\
    (yyyy-mm-dd)  &                  &(MHz)        &             & ($^\mathrm{h}$:$^\mathrm{m}$:$^\mathrm{s}$) & ($\degr$:$'$:$''$) & (mJy)              &(maj, min, PA)         \\ 
        (1)       & (2)              & (3)         & (4)         & (5)               & (6)              & (7)               &(8)                                                 \\ 
     \hline   
     2019-04-02   & VLA              &3000         &NE           &17:02:45.31        &13:01:02.3        &$7.7\pm0.3$      &$0.67''\times0.23''$, $122\degr$                      \\  
                  &                  &             &SW           &17:02:43.90        &13:00:51.9        &$2.2\pm0.5$      &$2.8''\times2.1''$, $26\degr$                         \\ 
     2021-10-02   & VLA              &3000         &NE           &17:02:45.32        &13:01:02.2        &$8.1\pm0.3$      &$1.3''\times0.44''$, $56\degr$                        \\
                  &                  &             &SW           &17:02:43.90        &13:00:51.9        &$3.0\pm0.4$      &$4.5''\times2.3''$, $20\degr$                         \\
    1995-02-27    & VLA              &1400         &NE+SW        &17:02:45.18        &13:01:01.0        &$27.8\pm0.4$     &$18.9''\times12.8''$, $15\degr$                       \\     
    2021-06-07    & GMRT             &1383         &NE           &17:02:45.11        &13 01 00.2        &$7.6\pm0.2$      &$1.2''\times0.88''$, $39\degr$                                 \\
                  &                  &             &SW           &17:02:43.68        &13 00 49.6        &$4.4\pm0.3$      &$<2.4''$                      \\
     2019-04-24   & ASKAP            &888          &NE           &17:02:45.29        &13:01:00.9        &$13.6\pm1.0$     &$10.6''\times0.80''$, $171\degr$                       \\
                  &                  &             &SW           &17:02:43.83        &13:00:51.6        &$5.3\pm0.9$      &$11.3''\times5.9''$, $164\degr$                       \\
     2020-05-01   & ASKAP            &888          &NE           &17:02:45.08        &13:01:05.0        &$12.2\pm0.7$     &$7.8''\times5.1''$, $112\degr$                        \\
                  &                  &             &SW           &17:02:43.66        &13:00:55.5        &$6.0\pm0.3$      &$7.4''\times6.2''$, $30\degr$                         \\
    2021-05-30    & GMRT             &607          &NE           &17:02:45.10        &13:01:00.3        &$10.1\pm0.2$     &$2.3'' \times 1.5''$, $129\degr$                      \\
                  &                  &             &SW           &17:02:43.68        &13:00:49.7        &$5.6\pm0.2$      &$2.7'' \times 1.4''$, $18\degr$                       \\
     2021-05-28   & GMRT             &321          &NE           &17:02:45.10        &13:01:01.2        &$10.5\pm0.5$     &$<5.0''$                                             \\
                  &                  &             &SW           &17:02:43.85        &13:00:51.1        &$7.7\pm0.5$      &$<9.7''$                      \\
     2018-03-13   & MWA              &216          &NE+SW        &17:02:46.94        &13:00:43.5        &$28.6\pm5.7$     &$<11.1''$                                             \\
     2018-03-13   & MWA              &185          &NE+SW        &17:02:45.27        &13:01:15.7        &$32.5\pm6.5$     &$<10.2''$                                             \\
     2018-03-13   & MWA              &154          &NE+SW        &17:02:46.73        &13:00:38.7        &$26.8\pm5.4$     &$<22.5''$                                             \\
    \hline
    \end{tabular}
    \label{tab:fitpars}
\end{table*}

\begin{table}
\caption{RATAN-600 measurements for SRGE~J170245.3+130104 in 2020--2022. Columns 1-2 are the averaged observing epoch (JD) and the date. Column 3 is the number of observations. Columns 4–6 are the flux densities and their uncertainties (mJy) at 11.2, 8.2 and 4.7 GHz, respectively. `-' indicates unsuccessful detection of flux density above $3\sigma$.} 
\centering
\begin{tabular}{crcccc}
\hline

JD  & yyyy.yy & $N_{\rm obs}$ & $S_{11.2}$ & $S_{8.2}$ & $S_{4.7}$ \\
    &     &    & (mJy)                  & (mJy)             &  (mJy)  \\
    \hline
(1) & (2) & (3) & (4) & (5) & (6)\\
\hline
2459196 & 2020.95 & 10 & -- & -- & $7\pm2$ \\
2459234 & 2021.05 & 31 & -- & $9\pm2$ & $5\pm2$ \\
2459301 & 2021.24 & 3  & -- & -- & $5\pm3$ \\ 
2459325 & 2021.34 & 17 & -- & $10\pm2$ & $6\pm2$ \\
2459417 & 2021.54 & 27 & -- & $6\pm3$ & $5\pm3$ \\
2459496 & 2021.77 & 21 & $13\pm3$ & $12\pm3$ & $5\pm1$ \\
2459567 & 2021.96 & 8  & -- & -- & $8\pm2$ \\
2459606 & 2022.07 & 12 & -- & $10\pm2$ & $5\pm1$ \\
2459632 & 2022.15 & 6  & --      & -- & $13\pm3$ \\
2459684 & 2022.28 & 27 & $9\pm3$ & -- & $5\pm1$ \\
2459725 & 2022.33 & 14 & $23\pm7$ & $10\pm2$ & $7\pm2$ \\
2459753 & 2022.47 & 2  & -- &  -- & $10\pm2$ \\
2459765 & 2022.51 & 7  & -- & $10\pm2$ & $10\pm2$ \\
2459809 & 2022.63 & 6  & -- & $19\pm4$ & $15\pm3$ \\
2459861 & 2022.77 & 6  & $13\pm3$ & $7\pm2$ & $8\pm2$ \\
\hline
\end{tabular}
\label{tab:ratan}
\end{table} 

\section*{Acknowledgements}
This research has been supported by the National SKA Program of China (2022SKA0120102, 2018YFA0404603).
Y.-Q. L.  is supported by Shanghai Post-doctoral Excellence Program.
Y.Z. is sponsored by Shanghai Sailing Program under grant number 22YF1456100.
Y.S., A.M., M.M., and T.M. are supported in the framework of the national project ‘Science’ by the Ministry of Science and Higher Education of the Russian Federation under the contract 075-15-2020-778.
We thank the TAC of the GMRT for approving the DDT observations (proposal id: DDT179) and the staff of the GMRT that made these observations possible. GMRT is run by the National Centre for Radio Astrophysics of the Tata Institute of Fundamental Research.
The data processing made use of the computing resource of the China SKA Regional Centre prototype, funded by the Ministry of Science and Technology of China and the Chinese Academy of Sciences. 
We acknowledge the Wajarri Yamatji as the traditional owners of the Murchison Radio-astronomy Observatory site.
The Australian Square Kilometre Array Pathfinder is part of the Australia Telescope National Facility which is managed by CSIRO. 
Operation of ASKAP is funded by the Australian Government with support from the National Collaborative Research Infrastructure Strategy.
The National Radio Astronomy Observatory is a facility of the National Science Foundation operated under cooperative agreement by Associated Universities, Inc. CIRADA is funded by a grant from the Canada Foundation for Innovation 2017 Innovation Fund (Project 35999), as well as by the Provinces of Ontario, British Columbia, Alberta, Manitoba and Quebec.

\section*{DATA AVAILABILITY}
The archive data used in the paper can be found in individual data bases, links to which are given in the text.   
The calibrated visibility data underlying this article can be requested from the corresponding author.


\bibliography{ref.bib}

\begin{thebibliography}{}
\makeatletter
\relax
\def\mn@urlcharsother{\let\do\@makeother \do\$\do\&\do\#\do\^\do\_\do\%\do\~}
\def\mn@doi{\begingroup\mn@urlcharsother \@ifnextchar [ {\mn@doi@}
  {\mn@doi@[]}}
\def\mn@doi@[#1]#2{\def\@tempa{#1}\ifx\@tempa\@empty \href
  {http://dx.doi.org/#2} {doi:#2}\else \href {http://dx.doi.org/#2} {#1}\fi
  \endgroup}
\def\mn@eprint#1#2{\mn@eprint@#1:#2::\@nil}
\def\mn@eprint@arXiv#1{\href {http://arxiv.org/abs/#1} {{\tt arXiv:#1}}}
\def\mn@eprint@dblp#1{\href {http://dblp.uni-trier.de/rec/bibtex/#1.xml}
  {dblp:#1}}
\def\mn@eprint@#1:#2:#3:#4\@nil{\def\@tempa {#1}\def\@tempb {#2}\def\@tempc
  {#3}\ifx \@tempc \@empty \let \@tempc \@tempb \let \@tempb \@tempa \fi \ifx
  \@tempb \@empty \def\@tempb {arXiv}\fi \@ifundefined
  {mn@eprint@\@tempb}{\@tempb:\@tempc}{\expandafter \expandafter \csname
  mn@eprint@\@tempb\endcsname \expandafter{\@tempc}}}

\bibitem[\protect\citeauthoryear{{Abdurro'uf} et~al.,}{{Abdurro'uf}
  et~al.}{2022}]{2022ApJS..259...35A}
{Abdurro'uf} et~al., 2022, \mn@doi [\apjs] {10.3847/1538-4365/ac4414}, \href
  {https://ui.adsabs.harvard.edu/abs/2022ApJS..259...35A} {259, 35}

\bibitem[\protect\citeauthoryear{{An}, {Wu}  \& {Hong}}{{An}
  et~al.}{2019}]{2019NatAs...3.1030A}
{An} T.,  {Wu} X.-P.,   {Hong} X.,  2019, \mn@doi [Nature Astronomy]
  {10.1038/s41550-019-0943-4}, \href
  {https://ui.adsabs.harvard.edu/abs/2019NatAs...3.1030A} {3, 1030}

\bibitem[\protect\citeauthoryear{{An} et~al.,}{{An}
  et~al.}{2020}]{2020NatCo..11..143A}
{An} T.,  et~al., 2020, \mn@doi [Nature Communications]
  {10.1038/s41467-019-14093-2}, \href
  {https://ui.adsabs.harvard.edu/abs/2020NatCo..11..143A} {11, 143}

\bibitem[\protect\citeauthoryear{{An}, {Wu}, {Lao}, {Guo}, {Xu}, {Lv}, {Zhang}
  \& {Zhang}}{{An} et~al.}{2022a}]{2022SCPMA..6529501A}
{An} T.,  {Wu} X.,  {Lao} B.,  {Guo} S.,  {Xu} Z.,  {Lv} W.,  {Zhang} Y.,
  {Zhang} Z.,  2022a, \mn@doi [Science China Physics, Mechanics, and Astronomy]
  {10.1007/s11433-022-1981-8}, \href
  {https://ui.adsabs.harvard.edu/abs/2022SCPMA..6529501A} {65, 129501}

\bibitem[\protect\citeauthoryear{{An}, {Wang}, {Zhang}, {Aditya}, {Hong}  \&
  {Cui}}{{An} et~al.}{2022b}]{2022MNRAS.511.4572A}
{An} T.,  {Wang} A.,  {Zhang} Y.,  {Aditya} J.~N.~H.~S.,  {Hong} X.,   {Cui}
  L.,  2022b, \mn@doi [\mnras] {10.1093/mnras/stac205}, \href
  {https://ui.adsabs.harvard.edu/abs/2022MNRAS.511.4572A} {511, 4572}

\bibitem[\protect\citeauthoryear{{Ba{\~n}ados} et~al.,}{{Ba{\~n}ados}
  et~al.}{2015}]{2015ApJ...804..118B}
{Ba{\~n}ados} E.,  et~al., 2015, \mn@doi [\apj] {10.1088/0004-637X/804/2/118},
  \href {https://ui.adsabs.harvard.edu/abs/2015ApJ...804..118B} {804, 118}

\bibitem[\protect\citeauthoryear{{Ba{\~n}ados}, {Carilli}, {Walter}, {Momjian},
  {Decarli}, {Farina}, {Mazzucchelli}  \& {Venemans}}{{Ba{\~n}ados}
  et~al.}{2018}]{2018ApJ...861L..14B}
{Ba{\~n}ados} E.,  {Carilli} C.,  {Walter} F.,  {Momjian} E.,  {Decarli} R.,
  {Farina} E.~P.,  {Mazzucchelli} C.,   {Venemans} B.~P.,  2018, \mn@doi
  [\apjl] {10.3847/2041-8213/aac511}, \href
  {https://ui.adsabs.harvard.edu/abs/2018ApJ...861L..14B} {861, L14}

\bibitem[\protect\citeauthoryear{{Belladitta}, {Moretti}, {Caccianiga},
  {Ghisellini}, {Cicone}, {Sbarrato}, {Ighina}  \& {Pedani}}{{Belladitta}
  et~al.}{2019}]{2019A&A...629A..68B}
{Belladitta} S.,  {Moretti} A.,  {Caccianiga} A.,  {Ghisellini} G.,  {Cicone}
  C.,  {Sbarrato} T.,  {Ighina} L.,   {Pedani} M.,  2019, \mn@doi [\aap]
  {10.1051/0004-6361/201935965}, \href
  {https://ui.adsabs.harvard.edu/abs/2019A&A...629A..68B} {629, A68}

\bibitem[\protect\citeauthoryear{{Belladitta} et~al.,}{{Belladitta}
  et~al.}{2020}]{2020A&A...635L...7B}
{Belladitta} S.,  et~al., 2020, \mn@doi [\aap] {10.1051/0004-6361/201937395},
  \href {https://ui.adsabs.harvard.edu/abs/2020A&A...635L...7B} {635, L7}

\bibitem[\protect\citeauthoryear{{Brammer}, {van Dokkum}  \& {Coppi}}{{Brammer}
  et~al.}{2008}]{2008ApJ...686.1503B}
{Brammer} G.~B.,  {van Dokkum} P.~G.,   {Coppi} P.,  2008, \mn@doi [\apj]
  {10.1086/591786}, \href
  {https://ui.adsabs.harvard.edu/abs/2008ApJ...686.1503B} {686, 1503}

\bibitem[\protect\citeauthoryear{{Condon}, {Cotton}, {Greisen}, {Yin},
  {Perley}, {Taylor}  \& {Broderick}}{{Condon}
  et~al.}{1998}]{1998AJ....115.1693C}
{Condon} J.~J.,  {Cotton} W.~D.,  {Greisen} E.~W.,  {Yin} Q.~F.,  {Perley}
  R.~A.,  {Taylor} G.~B.,   {Broderick} J.~J.,  1998, \mn@doi [\aj]
  {10.1086/300337}, \href
  {https://ui.adsabs.harvard.edu/abs/1998AJ....115.1693C} {115, 1693}

\bibitem[\protect\citeauthoryear{{Conroy}, {Gunn}  \& {White}}{{Conroy}
  et~al.}{2009}]{2009ApJ...699..486C}
{Conroy} C.,  {Gunn} J.~E.,   {White} M.,  2009, \mn@doi [\apj]
  {10.1088/0004-637X/699/1/486}, \href
  {https://ui.adsabs.harvard.edu/abs/2009ApJ...699..486C} {699, 486}

\bibitem[\protect\citeauthoryear{{Cutri} et~al.,}{{Cutri}
  et~al.}{2021}]{2014yCat.2328....0C}
{Cutri} R.~M.,  et~al., 2021, VizieR Online Data Catalog, \href
  {https://ui.adsabs.harvard.edu/abs/2014yCat.2328....0C} {p. II/328}

\bibitem[\protect\citeauthoryear{{De Rosa} et~al.,}{{De Rosa}
  et~al.}{2014}]{2014ApJ...790..145D}
{De Rosa} G.,  et~al., 2014, \mn@doi [\apj] {10.1088/0004-637X/790/2/145},
  \href {https://ui.adsabs.harvard.edu/abs/2014ApJ...790..145D} {790, 145}

\bibitem[\protect\citeauthoryear{{Dey} et~al.,}{{Dey}
  et~al.}{2019}]{2019AJ....157..168D}
{Dey} A.,  et~al., 2019, \mn@doi [\aj] {10.3847/1538-3881/ab089d}, \href
  {https://ui.adsabs.harvard.edu/abs/2019AJ....157..168D} {157, 168}

\bibitem[\protect\citeauthoryear{{Falcke}, {K{\"o}rding}  \& {Nagar}}{{Falcke}
  et~al.}{2004}]{2004NewAR..48.1157F}
{Falcke} H.,  {K{\"o}rding} E.,   {Nagar} N.~M.,  2004, \mn@doi [\nar]
  {10.1016/j.newar.2004.09.029}, \href
  {https://ui.adsabs.harvard.edu/abs/2004NewAR..48.1157F} {48, 1157}

\bibitem[\protect\citeauthoryear{{Frey}, {Paragi}, {Gurvits}, {Gab{\'a}nyi}  \&
  {Cseh}}{{Frey} et~al.}{2011}]{2011A&A...531L...5F}
{Frey} S.,  {Paragi} Z.,  {Gurvits} L.~I.,  {Gab{\'a}nyi} K.~{\'E}.,   {Cseh}
  D.,  2011, \mn@doi [\aap] {10.1051/0004-6361/201117341}, \href
  {https://ui.adsabs.harvard.edu/abs/2011A&A...531L...5F} {531, L5}

\bibitem[\protect\citeauthoryear{{Frey}, {Paragi}, {Fogasy}  \&
  {Gurvits}}{{Frey} et~al.}{2015}]{2015MNRAS.446.2921F}
{Frey} S.,  {Paragi} Z.,  {Fogasy} J.~O.,   {Gurvits} L.~I.,  2015, \mn@doi
  [\mnras] {10.1093/mnras/stu2294}, \href
  {https://ui.adsabs.harvard.edu/abs/2015MNRAS.446.2921F} {446, 2921}

\bibitem[\protect\citeauthoryear{{Ghisellini}, {Padovani}, {Celotti}  \&
  {Maraschi}}{{Ghisellini} et~al.}{1993}]{1993ApJ...407...65G}
{Ghisellini} G.,  {Padovani} P.,  {Celotti} A.,   {Maraschi} L.,  1993, \mn@doi
  [\apj] {10.1086/172493}, \href
  {https://ui.adsabs.harvard.edu/abs/1993ApJ...407...65G} {407, 65}

\bibitem[\protect\citeauthoryear{{Ghisellini}, {Haardt}, {Della Ceca},
  {Volonteri}  \& {Sbarrato}}{{Ghisellini} et~al.}{2013}]{2013MNRAS.432.2818G}
{Ghisellini} G.,  {Haardt} F.,  {Della Ceca} R.,  {Volonteri} M.,   {Sbarrato}
  T.,  2013, \mn@doi [\mnras] {10.1093/mnras/stt637}, \href
  {https://ui.adsabs.harvard.edu/abs/2013MNRAS.432.2818G} {432, 2818}

\bibitem[\protect\citeauthoryear{{Ghisellini}, {Haardt}, {Ciardi}, {Sbarrato},
  {Gallo}, {Tavecchio}  \& {Celotti}}{{Ghisellini}
  et~al.}{2015}]{2015MNRAS.452.3457G}
{Ghisellini} G.,  {Haardt} F.,  {Ciardi} B.,  {Sbarrato} T.,  {Gallo} E.,
  {Tavecchio} F.,   {Celotti} A.,  2015, \mn@doi [\mnras]
  {10.1093/mnras/stv1541}, \href
  {https://ui.adsabs.harvard.edu/abs/2015MNRAS.452.3457G} {452, 3457}

\bibitem[\protect\citeauthoryear{{Gordon} et~al.,}{{Gordon}
  et~al.}{2020}]{2020RNAAS...4..175G}
{Gordon} Y.~A.,  et~al., 2020, \mn@doi [Research Notes of the American
  Astronomical Society] {10.3847/2515-5172/abbe23}, \href
  {https://ui.adsabs.harvard.edu/abs/2020RNAAS...4..175G} {4, 175}

\bibitem[\protect\citeauthoryear{{Hotan} et~al.,}{{Hotan}
  et~al.}{2021}]{2021PASA...38....9H}
{Hotan} A.~W.,  et~al., 2021, \mn@doi [\pasa] {10.1017/pasa.2021.1}, \href
  {https://ui.adsabs.harvard.edu/abs/2021PASA...38....9H} {38, e009}

\bibitem[\protect\citeauthoryear{{Hurley-Walker} et~al.,}{{Hurley-Walker}
  et~al.}{2017}]{2017MNRAS.464.1146H}
{Hurley-Walker} N.,  et~al., 2017, \mn@doi [\mnras] {10.1093/mnras/stw2337},
  \href {https://ui.adsabs.harvard.edu/abs/2017MNRAS.464.1146H} {464, 1146}

\bibitem[\protect\citeauthoryear{{Hurley-Walker} et~al.,}{{Hurley-Walker}
  et~al.}{2022}]{2022PASA...39...35H}
{Hurley-Walker} N.,  et~al., 2022, \mn@doi [\pasa] {10.1017/pasa.2022.17},
  \href {https://ui.adsabs.harvard.edu/abs/2022PASA...39...35H} {39, e035}

\bibitem[\protect\citeauthoryear{{Ighina}, {Caccianiga}, {Moretti},
  {Belladitta}, {Della Ceca}, {Ballo}  \& {Dallacasa}}{{Ighina}
  et~al.}{2019}]{2019MNRAS.489.2732I}
{Ighina} L.,  {Caccianiga} A.,  {Moretti} A.,  {Belladitta} S.,  {Della Ceca}
  R.,  {Ballo} L.,   {Dallacasa} D.,  2019, \mn@doi [\mnras]
  {10.1093/mnras/stz2340}, \href
  {https://ui.adsabs.harvard.edu/abs/2019MNRAS.489.2732I} {489, 2732}

\bibitem[\protect\citeauthoryear{{Ighina}, {Belladitta}, {Caccianiga},
  {Broderick}, {Drouart}, {Moretti}  \& {Seymour}}{{Ighina}
  et~al.}{2021}]{2021A&A...647L..11I}
{Ighina} L.,  {Belladitta} S.,  {Caccianiga} A.,  {Broderick} J.~W.,  {Drouart}
  G.,  {Moretti} A.,   {Seymour} N.,  2021, \mn@doi [\aap]
  {10.1051/0004-6361/202140362}, \href
  {https://ui.adsabs.harvard.edu/abs/2021A&A...647L..11I} {647, L11}

\bibitem[\protect\citeauthoryear{{Intema}, {Jagannathan}, {Mooley}  \&
  {Frail}}{{Intema} et~al.}{2017}]{2017A&A...598A..78I}
{Intema} H.~T.,  {Jagannathan} P.,  {Mooley} K.~P.,   {Frail} D.~A.,  2017,
  \mn@doi [\aap] {10.1051/0004-6361/201628536}, \href
  {https://ui.adsabs.harvard.edu/abs/2017A&A...598A..78I} {598, A78}

\bibitem[\protect\citeauthoryear{{Jolley} \& {Kuncic}}{{Jolley} \&
  {Kuncic}}{2008}]{2008MNRAS.386..989J}
{Jolley} E.~J.~D.,  {Kuncic} Z.,  2008, \mn@doi [\mnras]
  {10.1111/j.1365-2966.2008.13082.x}, \href
  {https://ui.adsabs.harvard.edu/abs/2008MNRAS.386..989J} {386, 989}

\bibitem[\protect\citeauthoryear{{Kellermann}, {Sramek}, {Schmidt}, {Shaffer}
  \& {Green}}{{Kellermann} et~al.}{1989}]{1989AJ.....98.1195K}
{Kellermann} K.~I.,  {Sramek} R.,  {Schmidt} M.,  {Shaffer} D.~B.,   {Green}
  R.,  1989, \mn@doi [\aj] {10.1086/115207}, \href
  {https://ui.adsabs.harvard.edu/abs/1989AJ.....98.1195K} {98, 1195}

\bibitem[\protect\citeauthoryear{{Khorunzhev} et~al.,}{{Khorunzhev}
  et~al.}{2021}]{2021AstL...47..123K}
{Khorunzhev} G.~A.,  et~al., 2021, \mn@doi [Astronomy Letters]
  {10.1134/S1063773721030026}, \href
  {https://ui.adsabs.harvard.edu/abs/2021AstL...47..123K} {47, 123}

\bibitem[\protect\citeauthoryear{{Komatsu} et~al.,}{{Komatsu}
  et~al.}{2011}]{2011ApJS..192...18K}
{Komatsu} E.,  et~al., 2011, \mn@doi [\apjs] {10.1088/0067-0049/192/2/18},
  \href {https://ui.adsabs.harvard.edu/abs/2011ApJS..192...18K} {192, 18}

\bibitem[\protect\citeauthoryear{{Korolkov} \& {Pariiskii}}{{Korolkov} \&
  {Pariiskii}}{1979}]{1979S&T....57..324K}
{Korolkov} D.~V.,  {Pariiskii} I.~N.,  1979, \skytel, \href
  {https://ui.adsabs.harvard.edu/abs/1979S&T....57..324K} {57, 324}

\bibitem[\protect\citeauthoryear{{Lacy} et~al.,}{{Lacy}
  et~al.}{2020}]{2020PASP..132c5001L}
{Lacy} M.,  et~al., 2020, \mn@doi [\pasp] {10.1088/1538-3873/ab63eb}, \href
  {https://ui.adsabs.harvard.edu/abs/2020PASP..132c5001L} {132, 035001}

\bibitem[\protect\citeauthoryear{{Li} et~al.,}{{Li}
  et~al.}{2021}]{2021ApJ...906..135L}
{Li} J.-T.,  et~al., 2021, \mn@doi [\apj] {10.3847/1538-4357/abc750}, \href
  {https://ui.adsabs.harvard.edu/abs/2021ApJ...906..135L} {906, 135}

\bibitem[\protect\citeauthoryear{{Lobanov}}{{Lobanov}}{2005}]{2005astro.ph..3225L}
{Lobanov} A.~P.,  2005, \mn@doi [arXiv:astro-ph/050322]
  {arXiv:astro-ph/0503225}, \href
  {http://adsabs.harvard.edu/abs/2005astro.ph..3225L} {}

\bibitem[\protect\citeauthoryear{{Maraston}}{{Maraston}}{2005}]{2005MNRAS.362..799M}
{Maraston} C.,  2005, \mn@doi [\mnras] {10.1111/j.1365-2966.2005.09270.x},
  \href {https://ui.adsabs.harvard.edu/abs/2005MNRAS.362..799M} {362, 799}

\bibitem[\protect\citeauthoryear{{Mazzucchelli} et~al.,}{{Mazzucchelli}
  et~al.}{2017}]{2017ApJ...849...91M}
{Mazzucchelli} C.,  et~al., 2017, \mn@doi [\apj] {10.3847/1538-4357/aa9185},
  \href {https://ui.adsabs.harvard.edu/abs/2017ApJ...849...91M} {849, 91}

\bibitem[\protect\citeauthoryear{{McConnell} et~al.,}{{McConnell}
  et~al.}{2020}]{2020PASA...37...48M}
{McConnell} D.,  et~al., 2020, \mn@doi [\pasa] {10.1017/pasa.2020.41}, \href
  {https://ui.adsabs.harvard.edu/abs/2020PASA...37...48M} {37, e048}

\bibitem[\protect\citeauthoryear{{McMullin}, {Waters}, {Schiebel}, {Young}  \&
  {Golap}}{{McMullin} et~al.}{2007}]{2007ASPC..376..127M}
{McMullin} J.~P.,  {Waters} B.,  {Schiebel} D.,  {Young} W.,   {Golap} K.,
  2007, in {Shaw} R.~A.,  {Hill} F.,   {Bell} D.~J.,  eds,  Astronomical
  Society of the Pacific Conference Series Vol. 376, Astronomical Data Analysis
  Software and Systems XVI. p.~127

\bibitem[\protect\citeauthoryear{{Medvedev} et~al.,}{{Medvedev}
  et~al.}{2020}]{2020MNRAS.497.1842M}
{Medvedev} P.,  et~al., 2020, \mn@doi [\mnras] {10.1093/mnras/staa2051}, \href
  {https://ui.adsabs.harvard.edu/abs/2020MNRAS.497.1842M} {497, 1842}

\bibitem[\protect\citeauthoryear{{Miller}, {Brandt}, {Schneider}, {Gibson},
  {Steffen}  \& {Wu}}{{Miller} et~al.}{2011}]{2011ApJ...726...20M}
{Miller} B.~P.,  {Brandt} W.~N.,  {Schneider} D.~P.,  {Gibson} R.~R.,
  {Steffen} A.~T.,   {Wu} J.,  2011, \mn@doi [\apj]
  {10.1088/0004-637X/726/1/20}, \href
  {https://ui.adsabs.harvard.edu/abs/2011ApJ...726...20M} {726, 20}

\bibitem[\protect\citeauthoryear{{Momjian}, {Carilli}, {Ba{\~n}ados}, {Walter}
  \& {Venemans}}{{Momjian} et~al.}{2018}]{2018ApJ...861...86M}
{Momjian} E.,  {Carilli} C.~L.,  {Ba{\~n}ados} E.,  {Walter} F.,   {Venemans}
  B.~P.,  2018, \mn@doi [\apj] {10.3847/1538-4357/aac76f}, \href
  {https://ui.adsabs.harvard.edu/abs/2018ApJ...861...86M} {861, 86}

\bibitem[\protect\citeauthoryear{{Parijskij}}{{Parijskij}}{1993}]{1993IAPM...35....7P}
{Parijskij} Y.~N.,  1993, \mn@doi [IEEE Antennas and Propagation Magazine]
  {10.1109/74.229840}, \href
  {http://adsabs.harvard.edu/abs/1993IAPM...35....7P} {35, 7}

\bibitem[\protect\citeauthoryear{{Perger}, {Frey}, {Gab{\'a}nyi}  \&
  {T{\'o}th}}{{Perger} et~al.}{2017}]{2017FrASS...4....9P}
{Perger} K.,  {Frey} S.,  {Gab{\'a}nyi} K.~{\'E}.,   {T{\'o}th} L.~V.,  2017,
  \mn@doi [Frontiers in Astronomy and Space Sciences]
  {10.3389/fspas.2017.00009}, \href
  {https://ui.adsabs.harvard.edu/abs/2017FrASS...4....9P} {4, 9}

\bibitem[\protect\citeauthoryear{{Predehl} et~al.,}{{Predehl}
  et~al.}{2021}]{2021A&A...647A...1P}
{Predehl} P.,  et~al., 2021, \mn@doi [\aap] {10.1051/0004-6361/202039313},
  \href {https://ui.adsabs.harvard.edu/abs/2021A&A...647A...1P} {647, A1}

\bibitem[\protect\citeauthoryear{{Riseley}, {Scaife}, {Wise}  \&
  {Clarke}}{{Riseley} et~al.}{2017}]{2017A&A...597A..96R}
{Riseley} C.~J.,  {Scaife} A.~M.~M.,  {Wise} M.~W.,   {Clarke} A.~O.,  2017,
  \mn@doi [\aap] {10.1051/0004-6361/201629530}, \href
  {https://ui.adsabs.harvard.edu/abs/2017A&A...597A..96R} {597, A96}

\bibitem[\protect\citeauthoryear{{Rodr{\'\i}guez Zaur{\'\i}n}, {Tadhunter}  \&
  {Gonz{\'a}lez Delgado}}{{Rodr{\'\i}guez Zaur{\'\i}n}
  et~al.}{2008}]{2008MNRAS.384..875R}
{Rodr{\'\i}guez Zaur{\'\i}n} J.,  {Tadhunter} C.~N.,   {Gonz{\'a}lez Delgado}
  R.~M.,  2008, \mn@doi [\mnras] {10.1111/j.1365-2966.2007.12658.x}, \href
  {https://ui.adsabs.harvard.edu/abs/2008MNRAS.384..875R} {384, 875}

\bibitem[\protect\citeauthoryear{{Romani}, {Sowards-Emmerd}, {Greenhill}  \&
  {Michelson}}{{Romani} et~al.}{2004}]{2004ApJ...610L...9R}
{Romani} R.~W.,  {Sowards-Emmerd} D.,  {Greenhill} L.,   {Michelson} P.,  2004,
  \mn@doi [\apjl] {10.1086/423201}, \href
  {https://ui.adsabs.harvard.edu/abs/2004ApJ...610L...9R} {610, L9}

\bibitem[\protect\citeauthoryear{{Sbarrato} et~al.,}{{Sbarrato}
  et~al.}{2012}]{2012MNRAS.426L..91S}
{Sbarrato} T.,  et~al., 2012, \mn@doi [\mnras]
  {10.1111/j.1745-3933.2012.01332.x}, \href
  {https://ui.adsabs.harvard.edu/abs/2012MNRAS.426L..91S} {426, L91}

\bibitem[\protect\citeauthoryear{{Smith}, {Lonsdale}, {Lonsdale}  \&
  {Diamond}}{{Smith} et~al.}{1998}]{1998ApJ...493L..17S}
{Smith} H.~E.,  {Lonsdale} C.~J.,  {Lonsdale} C.~J.,   {Diamond} P.~J.,  1998,
  \mn@doi [\apjl] {10.1086/311122}, \href
  {https://ui.adsabs.harvard.edu/abs/1998ApJ...493L..17S} {493, L17}

\bibitem[\protect\citeauthoryear{{Sotnikova}}{{Sotnikova}}{2020}]{2020gbar.conf...32S}
{Sotnikova} Y.~V.,  2020, in {Romanyuk} I.~I.,  {Yakunin} I.~A.,  {Valeev}
  A.~F.,   {Kudryavtsev} D.~O.,  eds, Ground-Based Astronomy in Russia. 21st
  Century. pp 32--40, \mn@doi{10.26119/978-5-6045062-0-2_2020_32}

\bibitem[\protect\citeauthoryear{{Sotnikova} et~al.,}{{Sotnikova}
  et~al.}{2021}]{2021MNRAS.508.2798S}
{Sotnikova} Y.,  et~al., 2021, \mn@doi [\mnras] {10.1093/mnras/stab2114}, \href
  {https://ui.adsabs.harvard.edu/abs/2021MNRAS.508.2798S} {508, 2798}

\bibitem[\protect\citeauthoryear{{Spingola}, {Dallacasa}, {Belladitta},
  {Caccianiga}, {Giroletti}, {Moretti}  \& {Orienti}}{{Spingola}
  et~al.}{2020}]{2020A&A...643L..12S}
{Spingola} C.,  {Dallacasa} D.,  {Belladitta} S.,  {Caccianiga} A.,
  {Giroletti} M.,  {Moretti} A.,   {Orienti} M.,  2020, \mn@doi [\aap]
  {10.1051/0004-6361/202039458}, \href
  {https://ui.adsabs.harvard.edu/abs/2020A&A...643L..12S} {643, L12}

\bibitem[\protect\citeauthoryear{{Sunyaev} et~al.,}{{Sunyaev}
  et~al.}{2021}]{2021A&A...656A.132S}
{Sunyaev} R.,  et~al., 2021, \mn@doi [\aap] {10.1051/0004-6361/202141179},
  \href {https://ui.adsabs.harvard.edu/abs/2021A&A...656A.132S} {656, A132}

\bibitem[\protect\citeauthoryear{{Swarup}}{{Swarup}}{1991}]{1991ASPC...19..376S}
{Swarup} G.,  1991, in {Cornwell} T.~J.,  {Perley} R.~A.,  eds,  Astronomical
  Society of the Pacific Conference Series Vol. 19, IAU Colloq. 131: Radio
  Interferometry. Theory, Techniques, and Applications. pp 376--380

\bibitem[\protect\citeauthoryear{{Tingay} et~al.,}{{Tingay}
  et~al.}{2013}]{2013PASA...30....7T}
{Tingay} S.~J.,  et~al., 2013, \mn@doi [\pasa] {10.1017/pasa.2012.007}, \href
  {https://ui.adsabs.harvard.edu/abs/2013PASA...30....7T} {30, e007}

\bibitem[\protect\citeauthoryear{{Udovitskiy}, {Sotnikova}, {Mingaliev},
  {Tsybulev}, {Zhekanis}  \& {Nizhelskij}}{{Udovitskiy}
  et~al.}{2016}]{2016AstBu..71..496U}
{Udovitskiy} R.~Y.,  {Sotnikova} Y.~V.,  {Mingaliev} M.~G.,  {Tsybulev} P.~G.,
  {Zhekanis} G.~V.,   {Nizhelskij} N.~A.,  2016, \mn@doi [Astrophysical
  Bulletin] {10.1134/S1990341316040131}, \href
  {https://ui.adsabs.harvard.edu/abs/2016AstBu..71..496U} {71, 496}

\bibitem[\protect\citeauthoryear{{Vaughan}, {Edelson}, {Warwick}  \&
  {Uttley}}{{Vaughan} et~al.}{2003}]{2003MNRAS.345.1271V}
{Vaughan} S.,  {Edelson} R.,  {Warwick} R.~S.,   {Uttley} P.,  2003, \mn@doi
  [\mnras] {10.1046/j.1365-2966.2003.07042.x}, \href
  {https://ui.adsabs.harvard.edu/abs/2003MNRAS.345.1271V} {345, 1271}

\bibitem[\protect\citeauthoryear{{Verkhodanov}}{{Verkhodanov}}{1997}]{1997ASPC..125...46V}
{Verkhodanov} O.~V.,  1997, in {Hunt} G.,  {Payne} H.,  eds,  Astronomical
  Society of the Pacific Conference Series Vol. 125, Astronomical Data Analysis
  Software and Systems VI. p.~46

\bibitem[\protect\citeauthoryear{{Vito} et~al.,}{{Vito}
  et~al.}{2019}]{2019A&A...630A.118V}
{Vito} F.,  et~al., 2019, \mn@doi [\aap] {10.1051/0004-6361/201936217}, \href
  {https://ui.adsabs.harvard.edu/abs/2019A&A...630A.118V} {630, A118}

\bibitem[\protect\citeauthoryear{{Wang} et~al.,}{{Wang}
  et~al.}{2019}]{2019ApJ...884...30W}
{Wang} F.,  et~al., 2019, \mn@doi [\apj] {10.3847/1538-4357/ab2be5}, \href
  {https://ui.adsabs.harvard.edu/abs/2019ApJ...884...30W} {884, 30}

\bibitem[\protect\citeauthoryear{{Wayth} et~al.,}{{Wayth}
  et~al.}{2015}]{2015PASA...32...25W}
{Wayth} R.~B.,  et~al., 2015, \mn@doi [\pasa] {10.1017/pasa.2015.26}, \href
  {https://ui.adsabs.harvard.edu/abs/2015PASA...32...25W} {32, e025}

\bibitem[\protect\citeauthoryear{{Wayth} et~al.,}{{Wayth}
  et~al.}{2018}]{2018PASA...35...33W}
{Wayth} R.~B.,  et~al., 2018, \mn@doi [\pasa] {10.1017/pasa.2018.37}, \href
  {https://ui.adsabs.harvard.edu/abs/2018PASA...35...33W} {35, e033}

\bibitem[\protect\citeauthoryear{{Wright} et~al.,}{{Wright}
  et~al.}{2010}]{Wright2010}
{Wright} E.~L.,  et~al., 2010, \mn@doi [\aj] {10.1088/0004-6256/140/6/1868},
  \href {https://ui.adsabs.harvard.edu/abs/2010AJ....140.1868W} {140, 1868}

\bibitem[\protect\citeauthoryear{{Wu}, {Brandt}, {Miller}, {Garmire},
  {Schneider}  \& {Vignali}}{{Wu} et~al.}{2013}]{2013ApJ...763..109W}
{Wu} J.,  {Brandt} W.~N.,  {Miller} B.~P.,  {Garmire} G.~P.,  {Schneider}
  D.~P.,   {Vignali} C.,  2013, \mn@doi [\apj] {10.1088/0004-637X/763/2/109},
  \href {https://ui.adsabs.harvard.edu/abs/2013ApJ...763..109W} {763, 109}

\bibitem[\protect\citeauthoryear{{Zhang} et~al.,}{{Zhang}
  et~al.}{2017}]{2017MNRAS.468...69Z}
{Zhang} Y.,  et~al., 2017, \mn@doi [\mnras] {10.1093/mnras/stx392}, \href
  {https://ui.adsabs.harvard.edu/abs/2017MNRAS.468...69Z} {468, 69}

\bibitem[\protect\citeauthoryear{{Zhu}, {Brandt}, {Wu}, {Garmire}  \&
  {Miller}}{{Zhu} et~al.}{2019}]{2019MNRAS.482.2016Z}
{Zhu} S.~F.,  {Brandt} W.~N.,  {Wu} J.,  {Garmire} G.~P.,   {Miller} B.~P.,
  2019, \mn@doi [\mnras] {10.1093/mnras/sty2832}, \href
  {https://ui.adsabs.harvard.edu/abs/2019MNRAS.482.2016Z} {482, 2016}

\makeatother
\end{thebibliography}
\bibliographystyle{mnras}

\newpage
\onecolumn

\begin{table}
    \centering 
    \caption{Observation logs}
\begin{sideways}
\begin{tabular}{cccccccc} \hline\hline
    
Obs-date            &Telescope         &Frequency     &Bandwidth       &On-source time      &Beam                                          &$\sigma_{\rm rms}$      &$S_{\rm peak}$                 \\                   
(yyyy-mm-dd)        &  (Project Code)  &(MHz)         &(MHz)           &(min)               &(maj, min, PA)                                &(mJy beam$^{-1}$ )      &(mJy beam$^{-1}$)             \\    
\hline
2021-05-28          &GMRT(DDT179)     &321           &32               &30                  &$16.1''\times 8.4''$, $75\degr$               &0.25                    & 10.6                           \\
2021-05-30          &GMRT(DDT179)     &607           &32               &29                  &$8.1''\times 6.0''$, $51\degr$                &0.08                    & 9.3                           \\
2021-06-07          &GMRT(DDT179)     &1383          &32               &21                  &$4.2''\times 2.5''$, $70\degr$                &0.13                    &7.3                            \\
2019-04-24          &ASKAP(AS110)     &888           &288              &15                  &$20.3''\times10.9''$ , $-42\degr$             &0.38                    &12.2                           \\
2020-05-01       	&ASKAP(AS110)     &888           &288              &15                  &$14.7''\times13.0''$ , $-4\degr$              &0.26                    &10.6                           \\
1995-02-27          &VLA(NVSS)        &1400          &50               &                  &$45.0''\times45.0''$ , $0\degr$               &0.35                    &24                             \\
2019-04-02       	&VLA(VLASS1.2)    &3000          &2048             &                 &$2.8''\times2.4''$ , $-74\degr$               &0.17                    &7.1                            \\
2021-10-02          &VLA(VLASS2.2)    &3000          &2048             &                  &$3.3''\times2.3''$ , $55\degr$                &0.11                    &7.1                            \\
2018-03-12          &MWA(G0008)       &215.7         &31               &12                  &$63.3''\times47.6''$ ,$-31\degr$              &2.5                     &28.1                           \\
2018-07-07          &$\cdots$         & $\cdots$     & $\cdots$        &10                  & $\cdots$                                     & $\cdots$               & $\cdots$                      \\
2018-05-14          &$\cdots$         & $\cdots$     & $\cdots$        &12                  & $\cdots$                                     & $\cdots$               & $\cdots$                      \\
2018-05-21          &$\cdots$         & $\cdots$     & $\cdots$        &6                   & $\cdots$                                     & $\cdots$               & $\cdots$                      \\
2018-05-26          &$\cdots$         & $\cdots$     & $\cdots$        &18                  & $\cdots$                                     & $\cdots$               & $\cdots$                      \\
2019-05-21          &$\cdots$         & $\cdots$     & $\cdots$        &4                   & $\cdots$                                     & $\cdots$               & $\cdots$                      \\
2018-03-12          &$\cdots$         &185           &31               &12                  & $74.6''\times56.2''$ ,$-3\degr$              &                        &                               \\
2018-05-07          &$\cdots$         & $\cdots$     & $\cdots$        &14                  & $\cdots$                                     & $\cdots$               & $\cdots$                      \\
2018-05-14          &$\cdots$         & $\cdots$     & $\cdots$        &14                  & $\cdots$                                     & $\cdots$               & $\cdots$                      \\
2018-05-21          &$\cdots$         & $\cdots$     & $\cdots$        &12                  & $\cdots$                                     & $\cdots$               & $\cdots$                      \\       
2018-05-26          &$\cdots$         & $\cdots$     & $\cdots$        &14                  & $\cdots$                                     & $\cdots$               & $\cdots$                      \\         
2019-05-21          &$\cdots$         & $\cdots$     & $\cdots$        &12                  & $\cdots$                                     & $\cdots$               & $\cdots$                      \\  
2018-03-12          &$\cdots$         & 154          & 31              &14                  & $88.2''\times67.2''$, $-3\degr$              & $\cdots$               & $\cdots$                      \\
2018-05-07          &$\cdots$         & $\cdots$     & $\cdots$        &14                  & $\cdots$                                     & $\cdots$               & $\cdots$                      \\  
2018-05-14          &$\cdots$         & $\cdots$     & $\cdots$        &14                  & $\cdots$                                     & $\cdots$               & $\cdots$                      \\ 
2018-05-21          &$\cdots$         & $\cdots$     & $\cdots$        &14                  & $\cdots$                                     & $\cdots$               & $\cdots$                      \\          
2018-05-26          &$\cdots$         & $\cdots$     & $\cdots$        &14                  & $\cdots$                                     & $\cdots$               & $\cdots$                      \\              
2019-05-21          &$\cdots$         & $\cdots$     & $\cdots$        &12                  & $\cdots$                                     & $\cdots$               & $\cdots$                      \\  
2018-03-12          &$\cdots$         & 118          & 31              &10                  & $114''\times87''$, $-3\degr$                 & $\cdots$               & $\cdots$                      \\
2018-05-07          &$\cdots$         & $\cdots$     & $\cdots$        &14                  & $\cdots$                                     & $\cdots$               & $\cdots$                      \\  
2018-05-14          &$\cdots$         & $\cdots$     & $\cdots$        &14                  & $\cdots$                                     & $\cdots$               & $\cdots$                      \\ 
2018-05-21          &$\cdots$         & $\cdots$     & $\cdots$        &14                  & $\cdots$                                     & $\cdots$               & $\cdots$                      \\          
2018-05-26          &$\cdots$         & $\cdots$     & $\cdots$        &14                  & $\cdots$                                     & $\cdots$               & $\cdots$                      \\              
2019-05-21          &$\cdots$         & $\cdots$     & $\cdots$        &14                  & $\cdots$                                     & $\cdots$               & $\cdots$                      \\  
2018-03-12          &$\cdots$         & 88           & 31              &12                  & $154''\times118''$, $-3\degr$                & $\cdots$               & $\cdots$                      \\
2018-05-07          &$\cdots$         & $\cdots$     & $\cdots$        &12                  & $\cdots$                                     & $\cdots$               & $\cdots$                      \\  
2018-05-14          &$\cdots$         & $\cdots$     & $\cdots$        &14                  & $\cdots$                                     & $\cdots$               & $\cdots$                      \\ 
2018-05-21          &$\cdots$         & $\cdots$     & $\cdots$        &14                  & $\cdots$                                     & $\cdots$               & $\cdots$                      \\          
2018-05-26          &$\cdots$         & $\cdots$     & $\cdots$        &12                  & $\cdots$                                     & $\cdots$               & $\cdots$                      \\              
2019-05-21          &$\cdots$         & $\cdots$     & $\cdots$        &14                  & $\cdots$                                     & $\cdots$               & $\cdots$                      \\  
    \hline
    \end{tabular}
    \label{tab:obs}

\end{sideways}
\end{table}

\bsp	
\label{lastpage}
\end{document}